\documentclass[fleqn,usenatbib]{mnras}

\usepackage[T1]{fontenc}

\DeclareRobustCommand{\VAN}[3]{#2}
\let\VANthebibliography\thebibliography
\def\thebibliography{\DeclareRobustCommand{\VAN}[3]{##3}\VANthebibliography}

\usepackage{graphicx}	
\usepackage{amsmath}	
\usepackage{amssymb}	
\usepackage{float}
\usepackage[utf8]{inputenc}
\usepackage{hypcap}
\usepackage{xcolor}
\usepackage{tabularx}
\usepackage{footnote}
\usepackage{subfigure}

\newcommand{\gaia}{{\sf Gaia}}

\newcommand{\kepler}{{\it Kepler}}
\newcommand{\ktwo}{{\it K2}}
\newcommand{\degree}{$^{\circ}$}
\newcommand{\porb}{$P_{\rm orb}$}
\newcommand{\teff}{$T_{\rm eff}$}

\newcommand{\teffb}{$T_{\rm eff,2}$}
\newcommand{\logg}{$\log g$}
\newcommand{\vsini}{$v\sin i$}

\newcommand{\feh}{[Fe/H]}

\newcommand{\kms}{km\,s$^{-1}$}

\newcommand{\tara}{EPIC2753}
\newcommand{\tarb}{EPIC5147}

\title[Characterization of EPIC2753 and EPIC5147]{Optical characterization and Radial velocity monitoring with Belgian and Indian Telescopes (ORBIT): the eclipsing binaries EPIC\,211982753 and EPIC\,211915147}

\author[Panchal et al.]{
Alaxender Panchal,$^{1,2}$\thanks{E-mail: alaxender@aries.res.in}
Y. C. Joshi,$^{1}$
Peter De Cat,$^{3}$
Patricia Lampens,$^{3}$
Aruna Goswami,$^{4}$
S. N. Tiwari$^{2}$
\\
$^{1}$Aryabhatta Research Institute of observational sciencES (ARIES). Nainital, Uttrakhand, India\\
$^{2}$Department of Physics, DDU Gorakhpur University, Gorakhpur, India\\
$^{3}$Royal Observatory of Belgium, Ringlaan 3, B-1180 Brussels, Belgium\\
$^{4}$Indian Institute of Astrophysics, Bengaluru, Karnataka, India\\
}

\date{Accepted XXX. Received YYY; in original form ZZZ}

\pubyear{2022}

\begin{document}
\label{firstpage}
\pagerange{\pageref{firstpage}--\pageref{lastpage}}
\maketitle

\begin{abstract}
The \ktwo\ eclipsing binary candidates EPIC\,211982753 (hereinafter called EPIC2753) and EPIC\,211915147 (hereinafter called EPIC5147) are characterized with the help of photometric and high-resolution spectroscopic data. The light curve analysis uses the $R_{c}$-band photometric data from the 1.3-m Devasthal Fast Optical Telescope (DFOT, India), ASAS-3 and \ktwo\ observations. High-resolution \'echelle spectra are collected using the HERMES spectrograph at the 1.2-m MERCATOR telescope (La Palma, Spain). The synthetic light and radial velocity curves are generated with the help of the modeling package PHOEBE 1.0. The orbital period analysis based on the $\sim$3.2 years of \ktwo\ observations does not show any change in the orbital period of both targets. The component masses $M_{1,2}$ are estimated as 1.69(0.02) and 1.59(0.02) $M_{\odot}$ for EPIC2753, and 1.48(0.01) and 1.27(0.01) $M_{\odot}$ for EPIC5147. Both systems are high mass-ratio eclipsing binaries with q>0.85. The component radii $R_{1,2}$ are found to be 1.66(0.02) and 1.53(0.02) $R_{\odot}$ for EPIC2753, and 1.80(0.05) and 1.42(0.05) $R_{\odot}$ for EPIC5147. The distances of EPIC2753 and EPIC5147 are determined as 238(4) and 199(5) pc, respectively. MESA Isochrones and Stellar Tracks are used to understand the evolutionary status of both systems.
\end{abstract}


   
\section{Introduction} \label{sec:intro}
The stellar parameter determination is more precise and easier in the case of eclipsing binaries (EBs) as compared to single stars. Both photometric and spectroscopic observations are required for a complete orbital solution and direct parameter determination of EBs. In addition to the accurate 
fundamental parameters, EBs offer a way to explore the interaction between the components, evolution and formation of multiple systems, distance measurements, etc. Masses and radii determined using double-lined eclipsing binaries can be accurate up to 3$\%$ depending on the quality of photometric and spectroscopic data \citep{2010A&ARv..18...67T}. The information about luminosity ratio, radius ratio, inclination, orbital period (\porb) and eccentricity is derived from photometric time series. The multi-epoch radial velocity (RV) data give information about the mass ratio (q=$m_{2}/m_{1}$) of the system. The use of long-term photometric observations is required to detect any \porb\ variation signatures which can be used further to reveal underlying mechanisms for the observed \porb\ change. The \porb\ analysis is used to understand the processes of mass transfer, magnetic cycles as well as to infer the presence of any additional component in the system. There are thousands of EBs discovered by different ground- and space-based surveys during the last two decades \citep{2003ASPC..294..405S, 2010Sci...327..977B, 2015JATIS...1a4003R}. The high-precision and long-term continuous photometric observations by the space missions \kepler/\ktwo\ \citep{2010Sci...327..977B} and TESS \citep{2015JATIS...1a4003R} offer additional opportunities to study stellar activity, pulsation and \porb\ evolution of EB components. However RV data are unavailable for most of these systems.

On the basis of Roche lobe geometry, the EBs are divided into detached, semi-detached and contact systems. In detached binaries, radius of each component is smaller than Roche surface. In case of semi-detached systems, one of the components fills its Roche lobe whereas both components are found to fill or over-fill their Roche lobes in contact binaries. Due to the the large distance between the components of widely detached binaries, the components do not interact with each other during their evolution. The binary components in such systems  evolve independently of each other like an isolated star \citep{2002MNRAS.329..897H} and follow standard models of stellar evolution. Binaries are believed to be formed by fragmentation or from third-body capture  \citep{1975MNRAS.172P..15F, 2004A&A...414..633G, 2011psf..book.....B, 2017ApJS..230...15M}. Contact binary systems are formed from short-period detached or semi-detached systems via angular momentum loss \citep{1970PASJ...22..317O, 2002MNRAS.336..705B, 2020MNRAS.491.5158T}. The precise parameters of the components of EB systems can be used for testing stellar evolution models as well as to study multiple system evolution \citep{2017A&A...608A..62H}.

The systems EPIC2753 (= HD\,69735) and EPIC5147 (= HD\,73470) were both detected thanks to the STEREO mission of NASA. \citet{2011MNRAS.416.2477W} reported 163 EBs with brightness less than 10.5 mag. Out of these 163 EBs, 122 systems (including EPIC2753) were newly reported. \citet{2012MNRAS.427.2298W} analyzed the STEREO spacecraft data with a matched filter algorithm to detect low-amplitude signals. A total of nine low-mass EBs were detected while some other systems (including EPIC5147) were reported as potential follow-up sources due to their grazing eclipse features. Both systems are mentioned as EB candidates in \citet{2016AA...594A.100B}, who searched the \ktwo\ data (C1 to C6) for transit signals using a modified version of the CoRoT alarm pipeline. \citet{2016AA...594A.100B} reported 172 planetary and 327 EB candidates from \ktwo\ data for follow-up observations.
 
In the framework of the Belgo-Indian Network for Astronomy \& Astrophysics (BINA) project, we have initiated a long-term programme called "Optical characterization and Radial velocity monitoring with Belgian and Indian Telescopes {\it (ORBIT)}" which aims at collecting ground-based photometric and high-resolution spectroscopic observations of few selected low-mass eclipsing binary and exoplanet candidates to perform an in-depth characterization of their physical nature \citep{2019BSRSL..88...82J}. 
 
In this paper, the fundamental stellar parameters of the two systems are derived using light curves (LCs) and RV curves fitting. The systems have not been investigated in any previous studies. The basic information about the targets collected in different surveys is given in Table~\ref{tar_info}. The outline of the paper is as follows. The detailed observations concerning the targets are given in \textsection\,2. In \textsection\,3, we update the ephemeris of both systems. The radial velocities are derived in \textsection\,4. The modeling work has been carried out in \textsection\,5. Finally, we discuss the results in \textsection\,6.
%
\begin{table}
\caption{Basic information about EPIC2753 and EPIC5147 taken from different surveys.}
\label{tar_info}
\centering              
\scriptsize         
\begin{tabular}{p{.6in}p{.6in}p{.6in}p{1.3in}}
\hline
Parameter            & EPIC2753            & EPIC5147            & Reference \\
\hline
RA  (\degree)        & 124.7503            & 129.7887            & \citet{2016AA...594A.100B}       \\
DEC (\degree)        & +19.9760            & +18.9519            & \citet{2016AA...594A.100B}       \\
Period (days)        & 5.389631            & 1.811266            & \citet{2016AA...594A.100B}       \\
B (mag)              & 9.067 $\pm$ 0.014   & 9.255 $\pm$ 0.021   & \citet{2000AA...355L..27H}       \\
V (mag)              & 8.769 $\pm$ 0.014   & 8.815 $\pm$ 0.021   & \citet{2000AA...355L..27H}       \\
R (mag)              & 8.541               & 8.500               & \citet{2017yCat.2346....0B}      \\
I (mag)              & 8.428               & 8.327               & \citet{2017yCat.2346....0B}      \\
J (mag)              & 8.236 $\pm$ 0.027   & 8.082 $\pm$ 0.018   & \citet{2003yCat.2246....0C}      \\
H (mag)              & 8.180 $\pm$ 0.017   & 7.923 $\pm$ 0.029   & \citet{2003yCat.2246....0C}      \\
K (mag)              & 8.154 $\pm$ 0.018   & 7.893 $\pm$ 0.029   & \citet{2003yCat.2246....0C}      \\
L (mag)              & 8.135 $\pm$ 0.022   & 7.814 $\pm$ 0.027   & \citet{Cutri2012wise.rept....1C} \\
M (mag)              & 8.147 $\pm$ 0.019   & 7.851 $\pm$ 0.021   & \citet{Cutri2012wise.rept....1C} \\
N (mag)              & 8.167 $\pm$ 0.021   & 7.855 $\pm$ 0.025   & \citet{Cutri2012wise.rept....1C} \\
$A_{V}$ (mag)        & 0.116 $\pm$ 0.006   & 0.063 $\pm$ 0.003   & \citet{2011ApJ...737..103S}      \\
$\pi$ (mas)          & 4.033 $\pm$ 0.021   & 4.851 $\pm$ 0.020   & \citet{2021AA...649A...1G}   \\
\hline                  
\end{tabular}
\end{table}
%

\section{Observations}\label{Data}
\subsection{Photometry}\label{Photo}
The 1.3-m Devasthal Fast Optical Telescope (DFOT) was used for photometric follow-up observations. The $2k\times2k$ conventional back-illuminated CCD camera with a gain of $2e^{-} ADU^{-1}$ and a read-out noise of $7.5e^{-}$ was used during the observations. The field of view (FoV) of $\sim 18^{\arcmin}\times18^{\arcmin}$ provides multiple comparison stars in the vicinity of target stars. EPIC2753 was observed in the $R_{c}$ band during primary eclipse on three different nights. The system EPIC5147 was observed for $\sim$45 hours on different nights in the $R_{c}$ band covering both primary and secondary eclipse. The observation log for the targets using the 1.3-m DFOT is given in Table~\ref{log_phota}. The raw telescope images were cleaned with the help of standard IRAF\footnote{IRAF is distributed by the National Optical Astronomy Observatories, which is operated by the Association of Universities for Research in Astronomy, Inc. (AURA) under cooperative agreement with the National Science Foundation} routines. The AstroImageJ (AIJ) software package \citep{2017AJ....153...77C} was used to extract the target and comparison star fluxes from the science images using the technique of aperture photometry. For EPIC2753, TIC 14435258 and TIC 14433509 were used as comparison and check star, respectively. The field stars TIC 175233354 and TIC 175233313 were used as comparison and check star, respectively, for EPIC5147.

\begin{table}
\scriptsize
\caption{The 1.3\, m $R_{c}$-band observation log for EPIC2753 (upper block) and EPIC5147 (lower block).}
\centering
\label{log_phota}
\begin{tabular}{p{.45in}p{.45in}p{0.45in}p{0.40in}p{0.35in}p{0.05in}}
\hline
  Date of & Start BJD & End BJD   & Total      & Exposure    & Obs.  \\
  obs.    &           &           & frames     &             & time  \\
          & (2450000+)&(2450000+) &            & (sec)       & (hrs) \\
\hline                         
 20190110 & 8494.4059 & 8494.4970 & 239        & 20          & 2.19  \\
 20190428 & 8602.1411 & 8602.1744 & 198        & 05          & 0.80  \\
 20200319 & 8928.1715 & 8928.2522 & 499        & 05          & 1.94  \\
 20201230 & 9214.2692 & 9214.4177 & 1235       & 05          & 3.56  \\
 20210217 & 9263.0746 & 9263.1215 & 449        & 02          & 1.13  \\
 20210217 & 9263.2215 & 9263.2583 & 399        & 02          & 0.88  \\
 20220130 & 9610.1959 & 9610.4427 & 2759       & 02          & 5.92  \\
 20220201 & 9612.1165 & 9612.2528 & 1499       & 02          & 3.27  \\
 20220219 & 9630.0895 & 9630.3289 & 2639       & 02          & 5.75  \\
 20220301 & 9640.1629 & 9640.3570 & 505        & 02          & 4.66  \\
 20220307 & 9646.3134 & 9646.3616 & 470        & 02          & 1.16  \\
 20220310 & 9649.1266 & 9649.1467 & 209        & 02          & 0.48  \\
 20220318 & 9657.0656 & 9657.2849 & 2066       & 02          & 5.26  \\
 20220319 & 9658.2724 & 9658.2803 & 90         & 02          & 0.19  \\
 20220320 & 9659.0596 & 9659.3042 & 2729       & 02          & 5.87  \\
 20220414 & 9684.1458 & 9684.1658 & 175        & 02          & 0.48  \\
 20220415 & 9685.0783 & 9685.0987 & 175        & 02          & 0.49  \\
 20220416 & 9686.0855 & 9686.1109 & 280        & 02          & 0.61  \\
 20220417 & 9687.0903 & 9687.1027 & 140        & 02          & 0.30  \\
 & & & & & \\
\hline
 & & & & & \\
 20200213 & 8893.2185 & 8893.3898 & 1015       & 03          & 4.11  \\
 20200224 & 8904.2490 & 8904.3287 & 500        & 05          & 1.91  \\
 20211207 & 9556.2522 & 9556.5157 & 2100       & 03          & 6.32  \\
\hline                                                    
\end{tabular}       
\end{table}

The \kepler\ mission was launched by NASA in 2009 to detect exoplanets in the Cygnus-Lyra region of the Milky Way \citep{2010Sci...327..977B}. The mission observed $\sim$150,000 stars with a 30 and 1 minute cadence during the prime mission period up to November 2012. The \kepler\ spacecraft lost two reaction wheels by May 2013 which affected its pointing accuracy. After a redefinition of the goals, the second phase of mission, referred to as \ktwo\ , started in May 2014 and lasted for another four years. The \ktwo\ mission observed in total 21 campaign fields close to the ecliptic equator (C00, C01, C02,..., C19). All of them, except C00, were observed for approximately 80 days.  Both the \kepler\ and \ktwo\ observations were done through a spectral bandpass from 400 to 850\,nm. The data of the observed targets are available at the Barbara A. Mikulski Archive for Space Telescopes (MAST)\footnote{https://mast.stsci.edu/portal/Mashup/Clients/Mast/Portal.html} and the NASA Exoplanet Archive\footnote{https://exoplanetarchive.ipac.caltech.edu/}. EPIC2753 was observed by \ktwo\ for 74 and 49 days during C05 and C18, respectively. The system EPIC7147 was observed by \ktwo\ during the campaigns C05, C16 and C18 for almost 75, 80 and 51 days, respectively. The available photometric time series were corrected for the spacecraft's pointing errors with the EPIC\,Variability Extraction and Removal for Exoplanet Science Targets (EVEREST) pipeline \citep{2016AJ....152..100L}. The \ktwo\ observation log is given in Table~\ref{log_photb}.\\

Both the systems were observed by the All Sky Automated Survey (ASAS-3) from December 2002 to December 2009. The ASAS-3 is installed in Las Campanas Observatory and operated by Carnegie Institution of Washington \citep{2001ASPC..246...53P}. Most of the observations were done in V band. For EPIC2753, $\sim$ 475 data points were available and $\sim$ 400 data points had good quality according to given flags. Similarly, for EPIC5147, $\sim$ 450 data points were given good quality flag out of total 565 data points. The data points with large uncertainty and poor photometric quality were excluded during analysis. 
\begin{table}
\caption{The \ktwo\ observation log for EPIC2753 and EPIC5147.}
\centering
\label{log_photb}
\scriptsize
\begin{tabular}{p{.4in}p{.45in}p{.45in}p{0.30in}p{0.30in}p{0.4in}}
\hline
\ktwo\   &BJD$_{start}$&BJD$_{end}$& Data        & Exposure  & Reduction \\
Campaign & (2450000+)  & (2450000+)& points      & (seconds) & Pipeline  \\
\hline
EPIC2753 &             &           &             &           &           \\
 C05     &  7140.5502  & 7214.4109 &  3402       & 1800      & EVEREST   \\
 C18     &  8253.2213  & 8302.3801 &  2265       & 1800      & EVEREST   \\
 \hline
EPIC5147 &             &           &             &           &           \\
 C05     &  7139.6107  & 7214.4318 &  3663       & 1800      & EVEREST   \\
 C16     &  8095.4678  & 8175.0229 &  3894       & 1800      & EVEREST   \\
 C18     &  8251.5462  & 8302.4010 &  2490       & 1800      & EVEREST   \\
\hline                                                    
\end{tabular}       
\end{table}

\begin{figure*}
\begin{center}
\subfigure{\includegraphics[width=18cm,height=5cm]{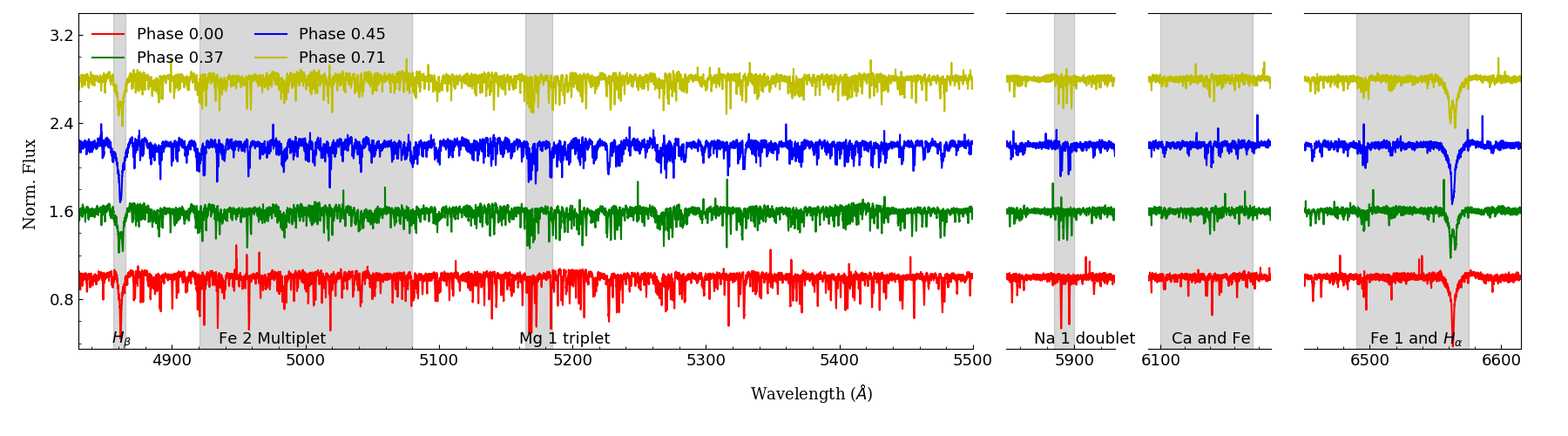}}\vspace{-15pt}
\subfigure{\includegraphics[width=18cm,height=5cm]{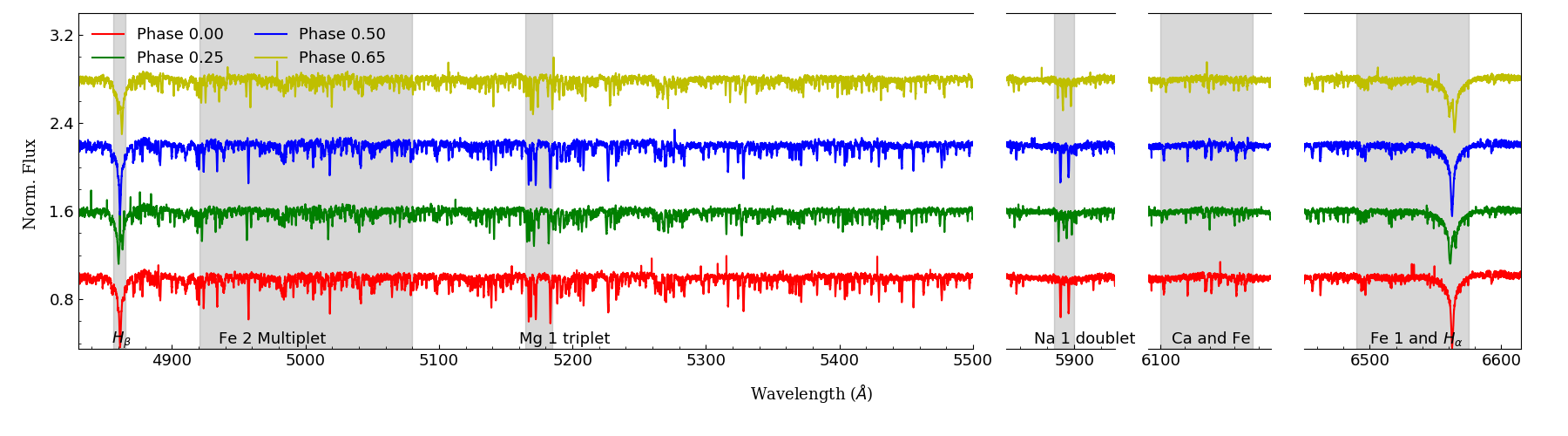}}
\caption{Sample spectra for EPIC2753 (upper) and EPIC5147 (lower) at different phases as observed using the HERMES spectrograph. Some spectral line groups are shown by shaded regions. The normalized spectra are vertically shifted for more clarity and the phases are given in upper left corner of each plot.}
\label{spec_obs}
\end{center}
\end{figure*}
%

\subsection{High Resolution Spectroscopy}\label{HRSpec}

The systems \tara\ and \tarb\ were observed using the High-Efficiency and high-Resolution Mercator Echelle Spectrograph (HERMES) mounted at the 1.2-m Mercator Telescope at the Roque de los Muchachos observatory on the Canary island La Palma in Spain \citep{Raskin2011A&A...526A..69R}. This high-resolution fibre-fed spectrograph provides a spectral coverage of 380-900 nm and a spectral resolution of $\sim$85\,000 in the high-resolution mode. A standard-silicon, thinned, back-illuminated 2048 $\times$ 4608 pixels CCD is used for recording the spectra. Both sources were observed in the simultaneous wavelength reference mode (HRF-WRF). Ten spectra were collected for EPIC2753 between 4 January to 7 February in 2019. For EPIC5147, 18 spectra were collected from 4 January to 26 February in 2019. The HERMES data reduction pipeline (DRS) V 5.0  was used for the reduction of the data. The normalisation was done with spline interpolation from the Python library SciPy. The overlapping region from adjacent orders was removed before combining all the orders in single spectra. Fig.~\ref{spec_obs} shows a selection of representative wavelength regions of the resulting normalised HERMES spectra of EPIC2753 and EPIC5147 at four different orbital phases.

\section{Updating Ephemeris}\label{orpe}

The \ktwo\ time series is used to obtain initial estimates of the \porb\ of each source. The photometric precision of the \ktwo\ data is better than that of the DFOT and ASAS-3 ground-based observations. Furthermore, continuous observations of the targets are available for many days (50-70 days) in the \ktwo\ survey. All the available data sets from different \ktwo\ campaigns are combined to determine the \porb. The \porb\ of each system is determined using the Period04 program. This program uses the discrete Fourier transform algorithm for the analysis of large time series with gaps \citep{2005CoAst.146...53L}. The \porb\ determination and its evolution in time are discussed in the following sections.

\subsection{EPIC2753}

The \porb\ of the system is determined as 5.390094 $\pm$ 0.000241 days using Period04. This value is in agreement with the 5.389631 days period reported by \cite{2016AA...594A.100B}. During the O-C analysis, we used HJD 2457141.732244 as the time of minimum brightness (TOM) at epoch 0. 
The orbital cycle numbers 
are calculated using the 5.390094 days period. The stability of \porb\ can be studied through an $(O-C)$-diagram. 
For this purpose, we calculated the TOMs from the \ktwo\ data using a quadratic (parabola) fitting in the minima regions of the light curve (LC).

\begin{figure}
\begin{center}
\subfigure{\includegraphics[width=8cm,height=6cm]{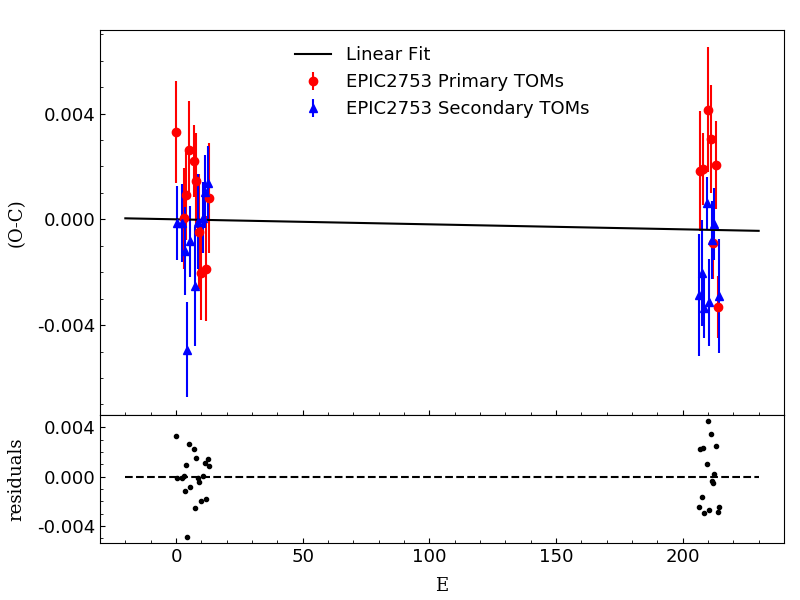}}\vspace{-18pt}
\subfigure{\includegraphics[width=8cm,height=6cm]{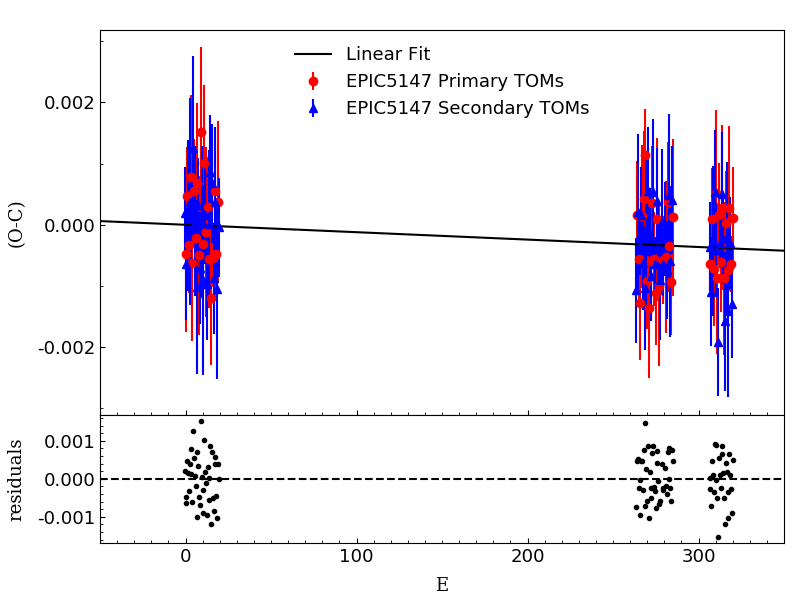}}
\caption{O-C diagram for EPIC2753 (upper) and EPIC5147 (lower) with linear fit.}
\label{oc_epic2}
\end{center}
\end{figure}
We thus determined 35 TOMs (17 primary TOMs + 18 secondary TOMs) for EPIC2753 from the \ktwo\ data. The updated linear ephemeris of the system can be described by the following equation: 
\begin{equation}
\label{li_epic1}
\begin{aligned}
HJD_{o} (E)=&2457141.7289(\pm 0.0005)\\
&+5.389918(\pm 0.000004)\times E
\end{aligned}
\end{equation} 
where, HJD$_{o}$(E) represents Heliocentric Julian date at the primary minimum of the orbital cycle number E. The upper panel of Fig.~\ref{oc_epic2} shows the best fit to E versus the O-C data with the residuals. The red and blue markers represent the primary and secondary minima, respectively. The O-C diagram does not show any variation and can be represented by the linear equation:
\begin{equation}
\label{li_epic_oc1}
\begin{aligned}
(O-C) =&-0.237836(\pm 508.62) \times10^{-6}\\
&-1.89206(\pm 3.68891) \times10^{-6}\times E
\end{aligned}
\end{equation}
Here, O-C is the difference between the observed and calculated $E^{th}$ TOM for the system in days. This fit is statistically equivalent with (O-C)=0 (as the errors on both variables are much larger than the values themselves), leading to the conclusion that the \porb\ is constant. The time basis available to us is only 200 cycles or ~3.2 yrs which is insufficient to reveal any period change caused by long term effects such as magnetic activity cycles or extra components. The information concerning TOMs, errors, orbital cycles, minima, O-C and their residuals is given in Table~\ref{OC_epicd}.

\begin{table}
\caption{Eclipse minima timings for EPIC2753 derived using Kepler/\ktwo\  data from campaign 05 and 18 observation.}
\label{OC_epicd}
\scriptsize
\begin{tabular}{p{.5in}p{.38in}p{0.3in}p{0.05in}p{0.5in}p{0.5in}}
\hline
HJD$_{\circ}$& Error     &Cycle   &Min&  O-C    & residuals \\
(2450000+)   &           &        &   &  (days)    & (days) \\
\hline
7141.732244 & 0.001928 & 000.0 & p &  0.003308 &  0.0033082 \\
7144.423768 & 0.001394 & 000.5 & s & -0.000128 & -0.0001268 \\
7155.203596 & 0.001485 & 002.5 & s & -0.000140 & -0.0001350 \\
7157.898731 & 0.001898 & 003.0 & p &  0.000035 &  0.0000409 \\
7160.592458 & 0.001668 & 003.5 & s & -0.001198 & -0.0011911 \\
7163.289521 & 0.001684 & 004.0 & p &  0.000905 &  0.0009128 \\
7165.978647 & 0.001798 & 004.5 & s & -0.004929 & -0.0049202 \\
7168.681164 & 0.001862 & 005.0 & p &  0.002628 &  0.0026377 \\
7171.372661 & 0.001336 & 005.5 & s & -0.000835 & -0.0008243 \\
7179.460596 & 0.001361 & 007.0 & p &  0.002220 &  0.0022335 \\
7182.150805 & 0.002250 & 007.5 & s & -0.002531 & -0.0025166 \\
7184.849762 & 0.001789 & 008.0 & p &  0.001466 &  0.0014814 \\
7187.543161 & 0.001797 & 008.5 & s & -0.000095 & -0.0000787 \\
7190.237754 & 0.002160 & 009.0 & p & -0.000462 & -0.0004447 \\
7195.626125 & 0.001812 & 010.0 & p & -0.002011 & -0.0019918 \\
7198.323152 & 0.001343 & 010.5 & s &  0.000056 &  0.0000761 \\
7203.714058 & 0.001381 & 011.5 & s &  0.001042 &  0.0010640 \\
7206.406114 & 0.001982 & 012.0 & p & -0.001862 & -0.0018391 \\
7209.104306 & 0.001406 & 012.5 & s &  0.001370 &  0.0013939 \\
7211.798695 & 0.002085 & 013.0 & p &  0.000799 &  0.0008238 \\
8254.744561 & 0.002299 & 206.5 & s & -0.002855 & -0.0024640 \\
8257.444191 & 0.002275 & 207.0 & p &  0.001815 &  0.0022069 \\
8260.135310 & 0.002012 & 207.5 & s & -0.002026 & -0.0016332 \\
8262.834198 & 0.001370 & 208.0 & p &  0.001902 &  0.0022958 \\
8265.523910 & 0.001143 & 208.5 & s & -0.003346 & -0.0029513 \\
8270.917781 & 0.000999 & 209.5 & s &  0.000605 &  0.0010016 \\
8273.616284 & 0.002348 & 210.0 & p &  0.004148 &  0.0045456 \\
8276.303955 & 0.001642 & 210.5 & s & -0.003141 & -0.0027425 \\
8279.005077 & 0.002043 & 211.0 & p &  0.003021 &  0.0034205 \\
8281.696244 & 0.001474 & 211.5 & s & -0.000772 & -0.0003716 \\
8284.391084 & 0.001375 & 212.0 & p & -0.000892 & -0.0004906 \\
8287.086759 & 0.001343 & 212.5 & s & -0.000177 &  0.0002253 \\
8289.783954 & 0.001663 & 213.0 & p &  0.002058 &  0.0024613 \\
8295.168500 & 0.001172 & 214.0 & p & -0.003316 & -0.0029109 \\
8297.863873 & 0.002144 & 214.5 & s & -0.002903 & -0.0024969 \\
\hline                  
\end{tabular}
\end{table}

\subsection{EPIC5147}

The \porb\ for the system is calculated as 3.62149 $\pm$ 0.00007 days using Period04, which is almost twice the \porb\ reported by \cite{2016AA...594A.100B}. The latter was derived using C05 data only as the other data sets (C16 and C18) were not available at that time. The orbital cycle numbers for the O-C analysis are calculated using the 3.62149 days period. We used HJD 2457142.197474 as the time of minimum brightness (TOM) at epoch 0.

Using the \ktwo\ data, we determined 111 TOMs (54 primary TOMs + 57 secondary TOMs) for the system. The updated linear ephemeris of the system can be described by the following straight line equation:
\begin{equation}
\label{li_epic2}
\begin{aligned}
HJD_{o}(E)=&2457142.1975(\pm 0.0001)\\
&+3.6215398(\pm 0.0000004)\times E
\end{aligned}
\end{equation} 
where, HJD$_{o}$ is Heliocentric Julian date at the primary minimum of the orbital cycle number E. The lower panel of Fig.~\ref{oc_epic2} shows the E versus O-C data for EPIC5147 with the best-fit line and the residuals. The red and blue markers represent the primary and secondary minima, respectively. The O-C diagram does not show any variation and can be represented by the linear equation:
\begin{equation}
\label{li_epic_oc2}
\begin{aligned}
(O-C) &=4.777081(\pm 1018.9) \times10^{-7} \\
&-1.21(\pm 0.44) \times10^{-7} \times E
\end{aligned}
\end{equation}
%
The analysis shows the stable nature of \porb\ over 3.2 years of observations. The information about TOMs, errors, orbital cycles, minima, O-C, and their residuals is given in Tables~\ref{OC_epica}, ~\ref{OC_epicb}, and ~\ref{OC_epicc}.
%
\begin{table}
\caption{Eclipse minima timings for EPIC5147 derived using Kepler/\ktwo\  data during campaign 5.}
\label{OC_epica}
\scriptsize
\begin{tabular}{p{.5in}p{.38in}p{0.3in}p{0.05in}p{0.5in}p{0.5in}}
\hline
HJD$_{\circ}$       & Error     &Cycle   &Min&  O-C    & residuals  \\
(2450000+)      &           &        &   &  (days) & (days)     \\
\hline
7140.3869020 & 0.0007434 & -000.5 & s & +0.0001985 & +0.0001974 \\
7142.1969896 & 0.0012666 & -000.0 & p & -0.0004844 & -0.0004849 \\
7144.0076015 & 0.0009135 & +000.5 & s & -0.0006430 & -0.0006429 \\
7145.8194884 & 0.0007902 & +001.0 & p & +0.0004734 & +0.0004741 \\
7147.6299309 & 0.0012335 & +001.5 & s & +0.0001454 & +0.0001467 \\
7149.4402279 & 0.0007784 & +002.0 & p & -0.0003281 & -0.0003262 \\
7151.2517057 & 0.0016897 & +002.5 & s & +0.0003792 & +0.0003817 \\
7153.0628811 & 0.0013343 & +003.0 & p & +0.0007841 & +0.0007873 \\
7154.8729775 & 0.0007374 & +003.5 & s & +0.0001100 & +0.0001138 \\
7156.6830177 & 0.0012744 & +004.0 & p & -0.0006203 & -0.0006159 \\
7158.4956594 & 0.0015121 & +004.5 & s & +0.0012509 & +0.0012559 \\
7160.3057294 & 0.0008565 & +005.0 & p & +0.0005504 & +0.0005560 \\
7162.1160139 & 0.0012270 & +005.5 & s & +0.0000644 & +0.0000706 \\
7163.9265102 & 0.0007920 & +006.0 & p & -0.0002098 & -0.0002030 \\
7165.7364772 & 0.0014260 & +006.5 & s & -0.0010133 & -0.0010059 \\
7167.5489511 & 0.0013074 & +007.0 & p & +0.0006901 & +0.0006981 \\
7169.3593553 & 0.0007659 & +007.5 & s & +0.0003238 & +0.0003324 \\
7171.1693048 & 0.0012976 & +008.0 & p & -0.0004972 & -0.0004880 \\
7172.9798649 & 0.0009096 & +008.5 & s & -0.0007076 & -0.0006978 \\
7174.7928658 & 0.0013748 & +009.0 & p & +0.0015228 & +0.0015332 \\
7176.6021609 & 0.0012434 & +009.5 & s & +0.0000474 & +0.0000584 \\
7178.4125730 & 0.0008047 & +010.0 & p & -0.0003110 & -0.0002994 \\
7180.2227451 & 0.0015406 & +010.5 & s & -0.0009094 & -0.0008972 \\
7182.0354321 & 0.0012814 & +011.0 & p & +0.0010071 & +0.0010199 \\
7183.8453559 & 0.0007990 & +011.5 & s & +0.0001604 & +0.0001738 \\
7185.6558483 & 0.0013923 & +012.0 & p & -0.0001177 & -0.0001037 \\
7187.4657641 & 0.0009107 & +012.5 & s & -0.0009724 & -0.0009578 \\
7189.2777945 & 0.0008149 & +013.0 & p & +0.0002875 & +0.0003028 \\
7191.0882841 & 0.0012178 & +013.5 & s & +0.0000066 & +0.0000225 \\
7192.8984857 & 0.0007961 & +014.0 & p & -0.0005623 & -0.0005458 \\
7194.7106738 & 0.0009439 & +014.5 & s & +0.0008553 & +0.0008724 \\
7196.5193851 & 0.0010885 & +015.0 & p & -0.0012039 & -0.0011862 \\
7198.3320358 & 0.0009687 & +015.5 & s & +0.0006763 & +0.0006946 \\
7200.1415943 & 0.0004790 & +016.0 & p & -0.0005357 & -0.0005168 \\
7201.9520256 & 0.0009038 & +016.5 & s & -0.0008749 & -0.0008554 \\
7203.7642171 & 0.0007583 & +017.0 & p & +0.0005461 & +0.0005662 \\
7205.5748171 & 0.0012270 & +017.5 & s & +0.0003756 & +0.0003963 \\
7207.3847410 & 0.0007734 & +018.0 & p & -0.0004710 & -0.0004497 \\
7209.1949384 & 0.0014773 & +018.5 & s & -0.0010441 & -0.0010222 \\
7211.0071249 & 0.0013266 & +019.0 & p & +0.0003719 & +0.0003944 \\
7212.8174863 & 0.0008083 & +019.5 & s & -0.0000372 & -0.0000141 \\
\hline             
\end{tabular}
\end{table}
%
\begin{table}
\caption{Same as Table~\ref{OC_epica} but for Kepler/\ktwo\  campaign 16.}
\label{OC_epicb}
\scriptsize
\begin{tabular}{p{.5in}p{.38in}p{0.3in}p{0.05in}p{0.5in}p{0.5in}}
\hline
HJD$_{\circ}$   & Error     &Cycle   &Min&  O-C    & residuals  \\
(2450000+)      &           &        &   &  (days)    & (days)  \\
\hline
8096.4724552 & 0.0008575 &  263.5 & s & -0.0010723 & -0.0007539 \\
8098.2844549 & 0.0008779 &  264.0 & p &  0.0001569 &  0.0004759 \\
8100.0952716 & 0.0012841 &  264.5 & s &  0.0002031 &  0.0005227 \\
8101.9052753 & 0.0007485 &  265.0 & p & -0.0005637 & -0.0002435 \\
8103.7162634 & 0.0004687 &  265.5 & s & -0.0003461 & -0.0000253 \\
8105.5260978 & 0.0009254 &  266.0 & p & -0.0012822 & -0.0009608 \\
8107.3383067 & 0.0007935 &  266.5 & s &  0.0001562 &  0.0004782 \\
8109.1490598 & 0.0011682 &  267.0 & p &  0.0001388 &  0.0004614 \\
8110.9590700 & 0.0007744 &  267.5 & s & -0.0006215 & -0.0002983 \\
8112.7708851 & 0.0011023 &  268.0 & p &  0.0004231 &  0.0007469 \\
8114.5801972 & 0.0010034 &  268.5 & s & -0.0010353 & -0.0007109 \\
8116.3931503 & 0.0007437 &  269.0 & p &  0.0011473 &  0.0014723 \\
8118.2026941 & 0.0012290 &  269.5 & s & -0.0000794 &  0.0002463 \\
8120.0126223 & 0.0007762 &  270.0 & p & -0.0009217 & -0.0005954 \\
8121.8248595 & 0.0010546 &  270.5 & s &  0.0005450 &  0.0008719 \\
8123.6337269 & 0.0011418 &  271.0 & p & -0.0013581 & -0.0010306 \\
8125.4457053 & 0.0007552 &  271.5 & s & -0.0001502 &  0.0001779 \\
8127.2560426 & 0.0004790 &  272.0 & p & -0.0005834 & -0.0002547 \\
8129.0665602 & 0.0007407 &  272.5 & s & -0.0008363 & -0.0005070 \\
8130.8785243 & 0.0009372 &  273.0 & p &  0.0003573 &  0.0006872 \\
8132.6894741 & 0.0011993 &  273.5 & s &  0.0005366 &  0.0008671 \\
8134.4991507 & 0.0007693 &  274.0 & p & -0.0005573 & -0.0002262 \\
8136.3098378 & 0.0004492 &  274.5 & s & -0.0006407 & -0.0003090 \\
8138.1201402 & 0.0008623 &  275.0 & p & -0.0011088 & -0.0007765 \\
8139.9324090 & 0.0008671 &  275.5 & s &  0.0003895 &  0.0007224 \\
8141.7428823 & 0.0013310 &  276.0 & p &  0.0000923 &  0.0004258 \\
8143.5531614 & 0.0007319 &  276.5 & s & -0.0003991 & -0.0000650 \\
8145.3633357 & 0.0013113 &  277.0 & p & -0.0009953 & -0.0006606 \\
8147.1741932 & 0.0009694 &  277.5 & s & -0.0009083 & -0.0005730 \\
8150.7966879 & 0.0011938 &  278.5 & s &  0.0000454 &  0.0003819 \\
8152.6068397 & 0.0007157 &  279.0 & p & -0.0005733 & -0.0002362 \\
8154.4175378 & 0.0005160 &  279.5 & s & -0.0006457 & -0.0003079 \\
8158.0396592 & 0.0007707 &  280.5 & s & -0.0000653 &  0.0002737 \\
8159.8499688 & 0.0012415 &  281.0 & p & -0.0005262 & -0.0001866 \\
8161.6605270 & 0.0008025 &  281.5 & s & -0.0007385 & -0.0003983 \\
8163.4724011 & 0.0009842 &  282.0 & p &  0.0003651 &  0.0007059 \\
8165.2832857 & 0.0013384 &  282.5 & s &  0.0004792 &  0.0008206 \\
8167.0932269 & 0.0007454 &  283.0 & p & -0.0003501 & -0.0000081 \\
8168.9037509 & 0.0012344 &  283.5 & s & -0.0005966 & -0.0002540 \\
8170.7141862 & 0.0008637 &  284.0 & p & -0.0009318 & -0.0005886 \\
8172.5262908 & 0.0008865 &  284.5 & s &  0.0004023 &  0.0007461 \\
8174.3367861 & 0.0012827 &  285.0 & p &  0.0001271 &  0.0004715 \\
\hline                   
\end{tabular}
\end{table}
\begin{table}
\caption{Same as Table~\ref{OC_epica} but for Kepler/\ktwo\  campaign 18.}
\label{OC_epicc}
\scriptsize
\begin{tabular}{p{.5in}p{.38in}p{0.3in}p{0.05in}p{0.5in}p{0.5in}}
\hline
HJD$_{\circ}$   & Error     &Cycle   &Min&  O-C    & residuals  \\
(2450000+)      &           &        &   &  (days)    & (days)  \\
\hline
8252.1994340 & 0.0007325 &  306.5 & s & -0.0003565 &  0.0000139 \\
8254.0099236 & 0.0004450 &  307.0 & p & -0.0006374 & -0.0002664 \\
8255.8202392 & 0.0008853 &  307.5 & s & -0.0010923 & -0.0007207 \\
8257.6321903 & 0.0008403 &  308.0 & p &  0.0000883 &  0.0004605 \\
8259.4426058 & 0.0012266 &  308.5 & s & -0.0002667 &  0.0001061 \\
8261.2529179 & 0.0009290 &  309.0 & p & -0.0007251 & -0.0003516 \\
8263.0649482 & 0.0010170 &  309.5 & s &  0.0005347 &  0.0009088 \\
8264.8757093 & 0.0013544 &  310.0 & p &  0.0005253 &  0.0009000 \\
8266.6855385 & 0.0007604 &  310.5 & s & -0.0004160 & -0.0000407 \\
8268.4958498 & 0.0012295 &  311.0 & p & -0.0008752 & -0.0004993 \\
8270.3055813 & 0.0008835 &  311.5 & s & -0.0019142 & -0.0015377 \\
8272.1184344 & 0.0008414 &  312.0 & p &  0.0001684 &  0.0005455 \\
8273.9287459 & 0.0004194 &  312.5 & s & -0.0002906 &  0.0000871 \\
8275.7391983 & 0.0008092 &  313.0 & p & -0.0006087 & -0.0002304 \\
8277.5510730 & 0.0010248 &  313.5 & s &  0.0004955 &  0.0008744 \\
8279.3616266 & 0.0013462 &  314.0 & p &  0.0002786 &  0.0006581 \\
8281.1718899 & 0.0007403 &  314.5 & s & -0.0002286 &  0.0001515 \\
8282.9820119 & 0.0012574 &  315.0 & p & -0.0008771 & -0.0004964 \\
8284.7920807 & 0.0011334 &  315.5 & s & -0.0015788 & -0.0011975 \\
8286.6044608 & 0.0008429 &  316.0 & p &  0.0000308 &  0.0004127 \\
8288.4150070 & 0.0012264 &  316.5 & s & -0.0001935 &  0.0001890 \\
8290.2252401 & 0.0007799 &  317.0 & p & -0.0007309 & -0.0003478 \\
8292.0353372 & 0.0014136 &  317.5 & s & -0.0014043 & -0.0010206 \\
8293.8477841 & 0.0013494 &  318.0 & p &  0.0002721 &  0.0006564 \\
8295.6579864 & 0.0007491 &  318.5 & s & -0.0002961 &  0.0000888 \\
8297.4684105 & 0.0004531 &  319.0 & p & -0.0006425 & -0.0002569 \\
8299.2785356 & 0.0008913 &  319.5 & s & -0.0012879 & -0.0009017 \\
8301.0907010 & 0.0008438 &  320.0 & p &  0.0001070 &  0.0004938 \\
\hline                  
\end{tabular}
\end{table}

\section{Radial Velocity Determination}\label{RVdet}

An accurate determination of the radial velocity requires appropriate templates for the EB components. Multiple synthetic spectra have been generated with the help of the stellar spectral synthesis program SPECTRUM \citep{1999ascl.soft10002G}. Synthetic stellar spectra are generated for effective temperature (\teff) ranging between 3500 and 10000\,K in steps of 250\,K, surface gravity (\logg) ranging between 3.0 and 5.0\,dex in steps of 0.5\,dex, and metallicity (\feh) ranging between -0.5 and 0.5\,dex in steps of 0.5\,dex. The ATLAS9 stellar atmospheric models \citep{2003IAUS..210P.A20C} are used during synthetic spectra generation\footnote{https://wwwuser.oats.inaf.it/castelli/grids.html}. The synthetic spectra are broadened using projected rotational velocity (\vsini) values up to 100\,\kms\, in steps of 5\,\kms. Target spectra observed closest to the primary and secondary eclipse are used to search the best RV template. The best templates are sorted out of 5985 synthetic templates on the basis of the shape of the cross-correlation function (CCF) computed with the IRAF FXCOR task \citep{2009arXiv0912.4755A}. The templates with high correlation height and Tonry $\&$ Davis R-value \citep{1979AJ.....84.1511T} are used as the final templates for RV determination. The templates whose temperatures were close to the \gaia\ DR3 temperatures showed a high correlation for primary components.

\begin{figure}
\begin{center}
\subfigure{\includegraphics[width=7cm,height=7cm]{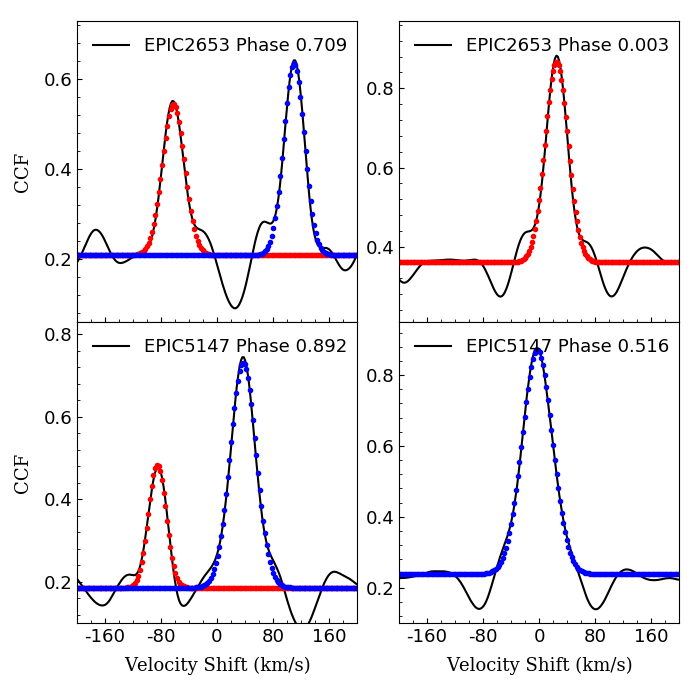}}
\caption{Some cross-correlation functions for EPIC2753 and EPIC5147 at phases close to quadratures and primary/secondary eclipses. The blue and red line are the fitted Gaussians for primary and secondary components, respectively.}
\label{epic_ccf}
\end{center}
\end{figure}

\begin{figure*}
\begin{center}
\subfigure{\includegraphics[width=18cm,height=6cm]{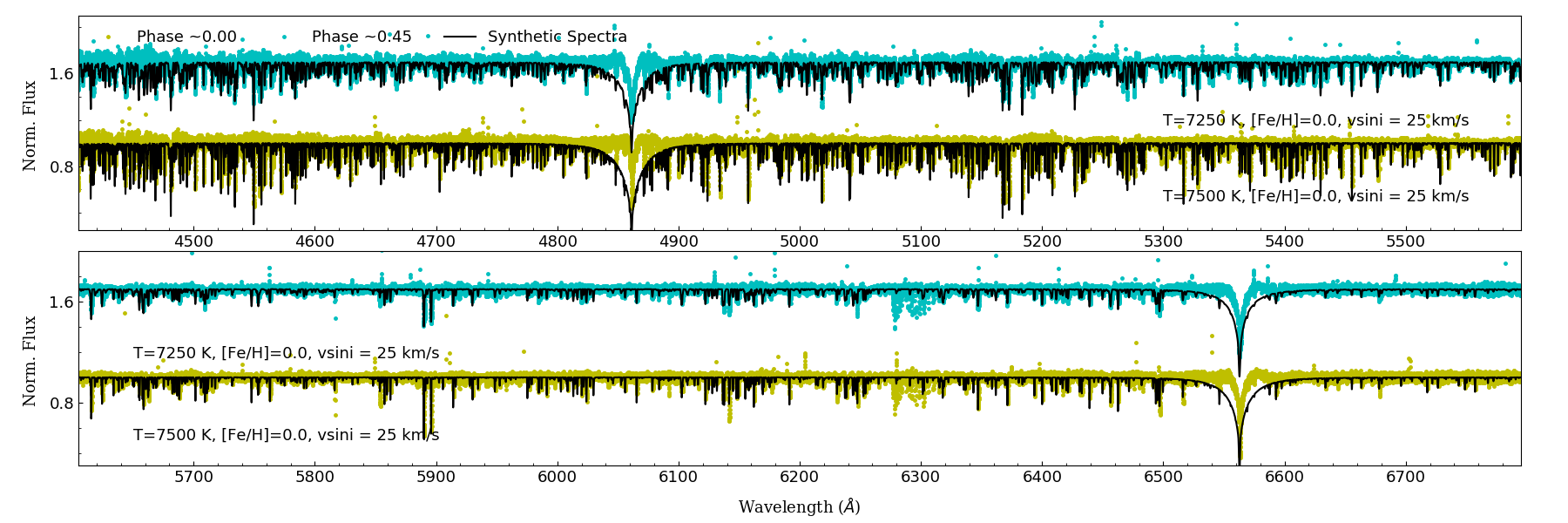}}\vspace{-19pt}
\subfigure{\includegraphics[width=18cm,height=6cm]{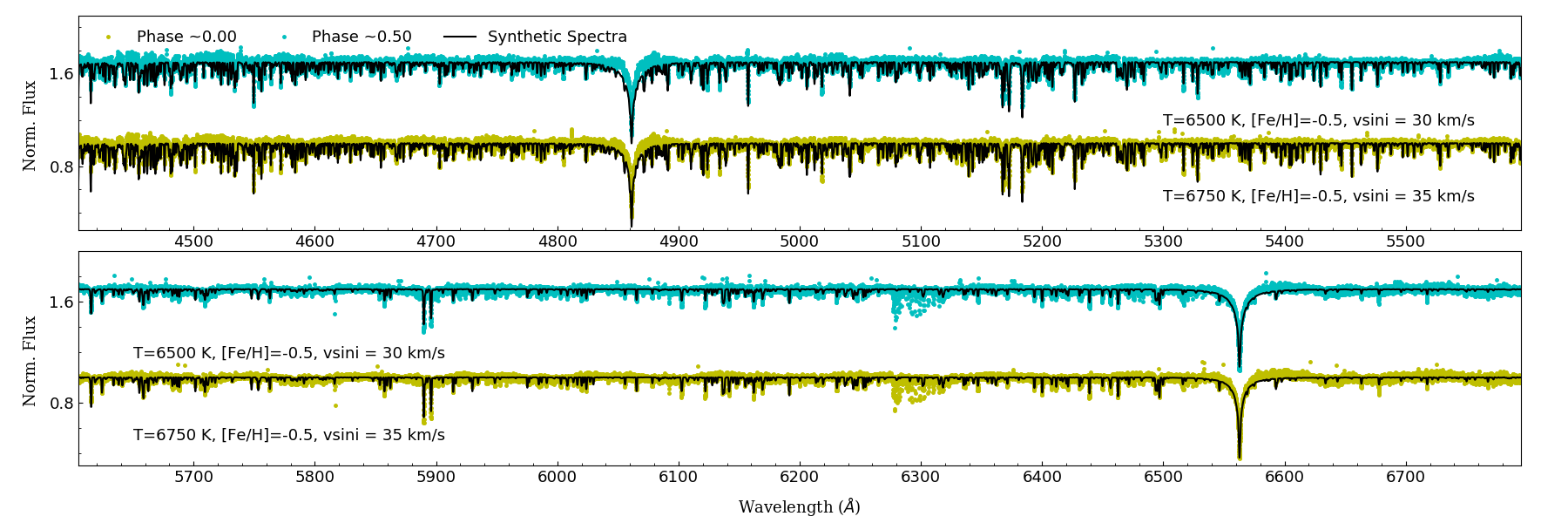}}
\caption{The observed HERMES spectra (in color) along with best fit synthetic spectra (in black) for EPIC2753 (upper panel) and EPIC5147 (lower panel) at an orbital phase around primary eclipse and secondary eclipse.}
\label{epic_sp}
\end{center}
\end{figure*}

 For EPIC2753, the primary component template (corresponding to the observed spectra close to the phase of secondary eclipse) is generated with \teff\ of 7500 K, \feh\ of 0.0, \logg\ of 4.0, and \vsini\ around 25\,\kms. The secondary component template (corresponding to the observed spectra close to the phase of primary eclipse) for EPIC2753 is generated using the \teff\ of 7250 K, \feh\ of 0.0, \logg\ of 4.0, and \vsini\ around 25\,\kms. For EPIC5147, the primary component template is generated with \teff\ of 6750 K, the \feh\ of -0.5, \logg\ of 4.0 and \vsini\ around 35\,\kms. The secondary component template for EPIC5147 is generated using the \teff\ of 6500 K, \feh\ of -0.5, \logg\ of 4.0, and \vsini\ around 30\,\kms. For the final determination of RVs, the template corresponding to a temperature of 7250 K is used for EPIC2753. For EPIC5147, the template corresponding to 6750 K is used to derive the RVs. The RVs are determined in five different wavelength regions excluding the hydrogen lines ($H_{\alpha}, H_{\beta}, H_{\gamma}$) and telluric lines. The wavelength range for these regions are 4225-4300, 4390-4500, 4505-4580, 4895-5200, and 5825-5950 {\AA}. The atmospheric parameters of the synthetic spectra may not be similar to the actual target atmospheric parameters as these are derived on the basis of CCF and use specific regions of spectra instead of complete spectra. The RV shifts are determined via Gaussian fitting to the FXCOR generated CCFs. In the CCFs for spectra close to primary/secondary eclipse, only one component could be detected due to blending of the spectral lines. Fig.~\ref{epic_ccf} shows some examples of CCFs for EPIC2753 and EPIC5147. Only one peak is visible for CCFs close to the phases of eclipses. The RV estimates from spectra close to eclipses are not used for computing the orbital solution. The calculated RVs along with their root mean square errors are given in Table~\ref{rv_ep}. The best synthetic spectra template selected via high CCF for both systems are shown in Fig.~\ref{epic_sp} along with the observed spectra.
 
%

%

\begin{table}
\caption{The RV data for both the components of EPIC2753 and EPIC5147.}             
\label{rv_ep}
\centering          
\begin{tabular}{c c r c r c}   
\hline\hline   
BJD       & Phase & $RV_{1}$&Error &$RV_{2}$ & Error  \\
(+2450000)&       & km/sec  &km/sec& km/sec  &        \\
          &       &         &      &         &        \\
\hline  

EPIC2753  &       &         &      &         &        \\
8487.6401 & 0.709 & +109.13 & 0.79 & -064.69 & 1.68  \\
8488.6142 & 0.890 & +081.49 & 0.56 & -035.52 & 1.41  \\
8489.6271 & 0.078 & -016.53 & 0.94 & +069.25 & 1.55  \\
8491.6412 & 0.451 & -000.23 & 0.77 & +051.64 & 1.26  \\
8492.6147 & 0.632 & +088.73 & 0.88 & -042.45 & 1.01  \\
8493.6189 & 0.818 & +104.86 & 0.59 & -059.57 & 1.08  \\
8494.6118 & 0.003 &   ---   & ---  & +024.60 & 0.68  \\
8495.6105 & 0.188 & -056.23 & 0.84 & +111.26 & 1.17  \\
8496.6003 & 0.372 & -037.78 & 0.68 & +091.24 & 1.07  \\
8521.5643 & 0.003 &   ---   & ---  & +024.84 & 0.67  \\
          &       &         &      &         &        \\
EPIC5147  &       &         &      &         &        \\
8487.6557 & 0.516 & +003.30 & 0.60 &   ---   &  ---   \\
8488.6281 & 0.784 & +087.61 & 0.92 & -100.11 & 3.03   \\
8489.6410 & 0.064 & -032.81 & 0.13 & +039.50 & 3.49   \\
8491.6550 & 0.620 & +061.59 & 0.46 & -069.14 & 0.75   \\
8492.6396 & 0.892 & +056.83 & 0.37 & -064.12 & 0.87   \\
8493.6318 & 0.166 & -075.39 & 0.38 & +090.25 & 0.31   \\
8494.6292 & 0.441 & -029.91 & 0.27 & +036.19 & 5.53   \\
8495.6281 & 0.717 & +087.69 & 0.85 & -099.88 & 4.72   \\
8496.6132 & 0.989 &   ---   & ---  & +002.41 & 0.40   \\
8520.5867 & 0.609 & +057.10 & 0.51 & -064.47 & 0.63   \\
8520.5988 & 0.612 & +058.38 & 0.50 & -065.63 & 0.54   \\
8522.5757 & 0.158 & -072.99 & 0.67 & +087.17 & 0.76   \\
8524.5703 & 0.709 & +086.42 & 0.58 & -098.79 & 3.84   \\
8525.5787 & 0.987 &   ---   & ---  & +003.10 & 0.53   \\
8526.5397 & 0.252 & -087.51 & 1.17 & +103.53 & 4.86   \\
8527.5523 & 0.532 & +019.31 & 0.27 & -018.70 & 3.70   \\
8528.5736 & 0.814 & +082.92 & 0.62 & -093.67 & 4.19   \\
8541.5097 & 0.386 & -057.30 & 0.39 & +068.36 & 1.54   \\
\hline                  
\end{tabular}
\vspace{1ex}
\end{table}

\section{Modeling}\label{Ana1}

\begin{figure*}
\begin{center}
\subfigure{\includegraphics[width=18cm,height=6cm]{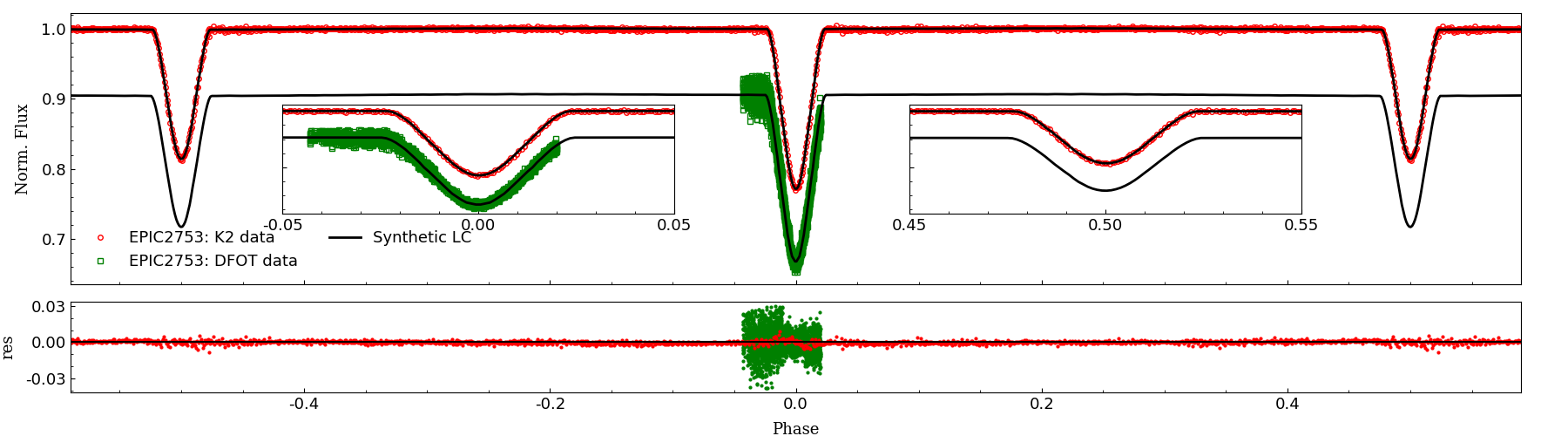}}
\caption{Combined \ktwo\ data from different campaigns and DFOT $R_{c}$-band data along with synthetic LCs (continuous line) for EPIC2753. }
\label{fit_epica}
\end{center}
\end{figure*}

\begin{figure*}
\begin{center}
\subfigure{\includegraphics[width=18cm,height=6cm]{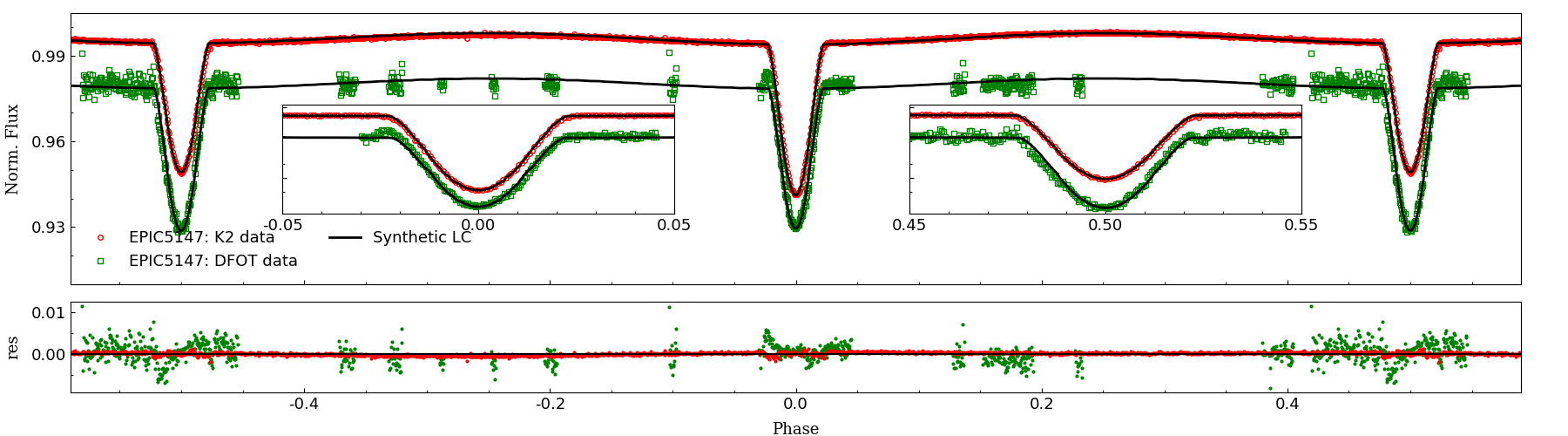}}
\caption{Same as Fig.~\ref{fit_epica} but for EPIC5147. }
\label{fit_epicb}
\end{center}
\end{figure*}

\begin{figure}
\begin{center}
\subfigure{\includegraphics[width=8cm,height=6cm]{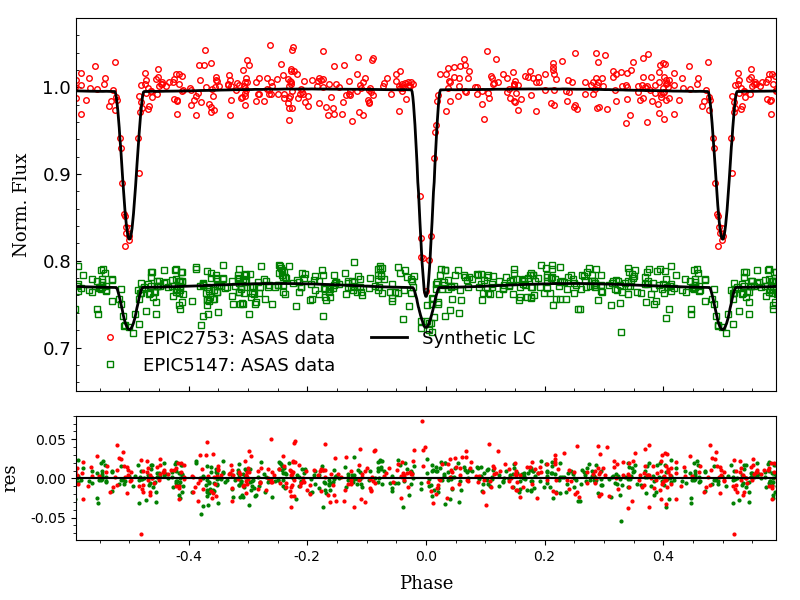}}
\caption{ASAS-3 observations for EPIC2753 and EPIC5147 with synthetic LCs. }
\label{fit_epicc}
\end{center}
\end{figure}

\begin{figure}
\begin{center}
\subfigure{\includegraphics[width=8cm,height=6cm]{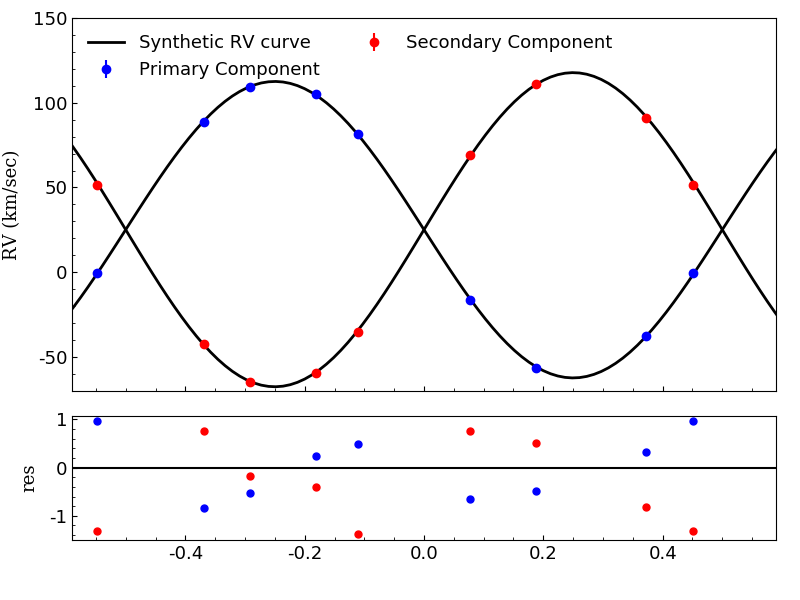}}\vspace{-15pt}
\subfigure{\includegraphics[width=8cm,height=6cm]{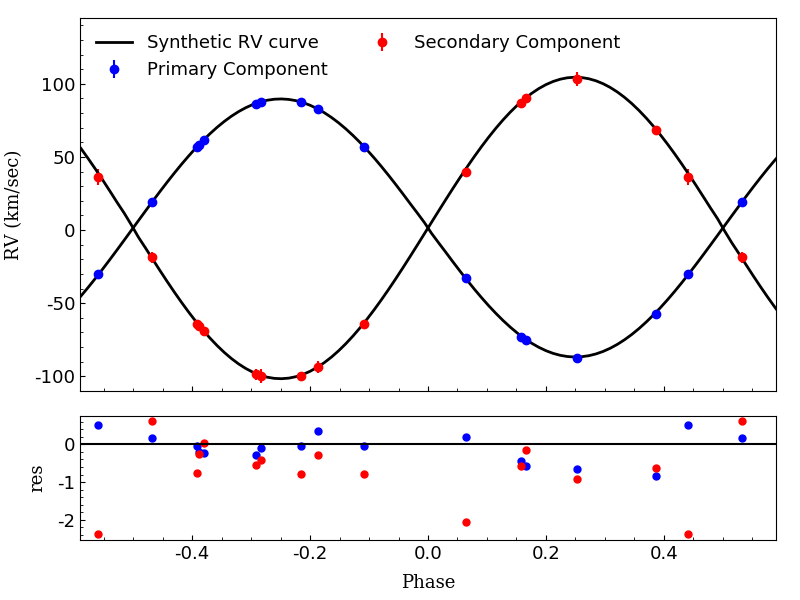}}
\caption{The RV variation for primary and secondary components of EPIC2753 (upper panel) and EPIC5147 (lower panel) with the model fit.}
\label{fit_epicd}
\end{center}
\end{figure}

For photometric and radial velocity data modeling, the python-based modeling package PHOEBE 1.0 (PHysics Of Eclipsing BinariEs) is used \citep{2005ApJ...628..426P}. The software is based on the popular FORTRAN-based WD program \citep{1971ApJ...166..605W}. The graphical user interface (GUI) in PHOEBE is used for obtaining initial estimates of the parameters by viewing the synthetic fit at regular intervals after multiple iterations. The PHOEBE scripter is used to refine the parameters and determine the uncertainties in the parameters. The radial velocity (RV) and the photometric data can be analyzed together. The TOM at $0^{th}$ epoch and \porb\ are used from section~\ref{orpe}. The temperature estimates available in \cite{2022yCat.1355....0G} are used as the $T_{\rm eff}$ of the primary component for both the sources. For EPIC2753 and EPIC5147, the \gaia\ DR-3 $T_{\rm eff}$ are reported as 7416($\pm$12) and 6654($\pm$ 19), respectively. The $T_{\rm eff}$ are determined by General Stellar Parametrizer from Photometry (GSP-Phot) pipeline using low-resolution BP/RP spectra. GSP-Phot uses an Aeneas algorithm which employs MCMC to optimize the parameters. The components with $T_{\rm eff}$ below 7200 K are assumed to have a convective envelope. The surface albedo (A; fraction of light reflected by a star) and gravity brightening (g; variation in the pole to equator surface brightness of a star due to its rotation) for convective envelope stars are taken as 0.5 and 0.32, respectively. The surface albedo (A) and gravity brightening (g) for radiative envelope stars ($T_{\rm eff} >$ 7200 K) are taken as 1.0 and 1.0, respectively. The limb darkening coefficients are updated by the software after each iteration using the tables by \cite{1993AJ....106.2096V}. Known parameters such as HJD$_{o}$, the \porb\, and the primary component's effective temperature are fixed during analysis. During the first run, both the photometric and RV data are modeled using the PHOEBE GUI, to estimate the remaining free parameters. Based on the shape of LCs, the "Detached model" is used for both the systems. The semi-major axis (a; separation between EB components), center of mass velocity ($V_{\gamma}$; velocity of the center of mass of the EB system), mass ratio (q=$m_{2}/m_{1}$), inclination (i; angle between EB orbital plane and the sky plane), secondary component's $T_{\rm eff}$, primary component passband luminosity ($l_{1}$; used to calculate the luminosity ratio for the components from input uncalibrated/calibrated LCs, not necessarily in standard units), primary and secondary component surface potential ($\Omega_{1/2}$; dimensionless potential or the modified Kopal potential) are set free. The eccentricity of both systems are taken as 0. The fitted LC and RV curves for each system are shown in Fig.~\ref{fit_epica}, Fig.~\ref{fit_epicb}, Fig.~\ref{fit_epicc}, and Fig.~\ref{fit_epicd}.

%

%

%

\begin{figure}
\begin{center}
\subfigure{\includegraphics[width=8cm,height=6cm]{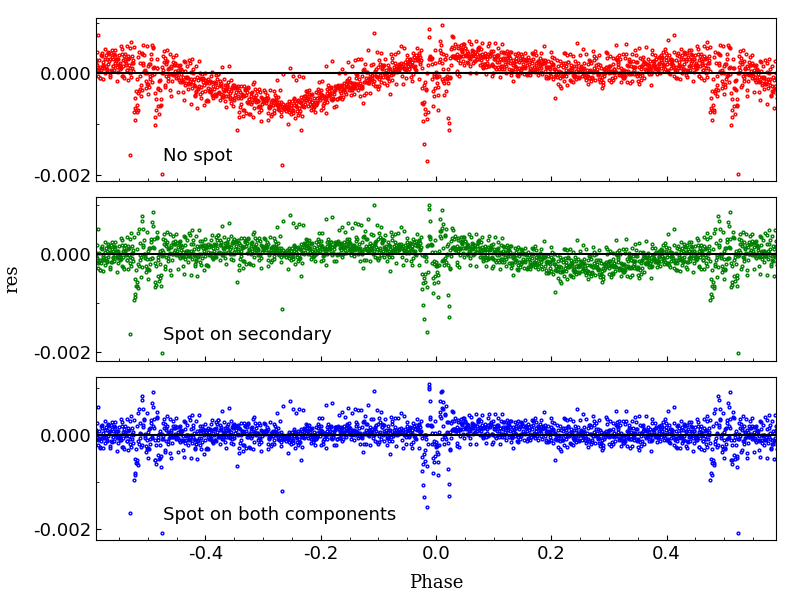}}
\caption{Change in the fit residuals for EPIC5147 after including spots.}
\label{fit_res}
\end{center}
\end{figure}

\begin{figure*}
\begin{center}
\subfigure{\includegraphics[width=15cm,height=6.5cm]{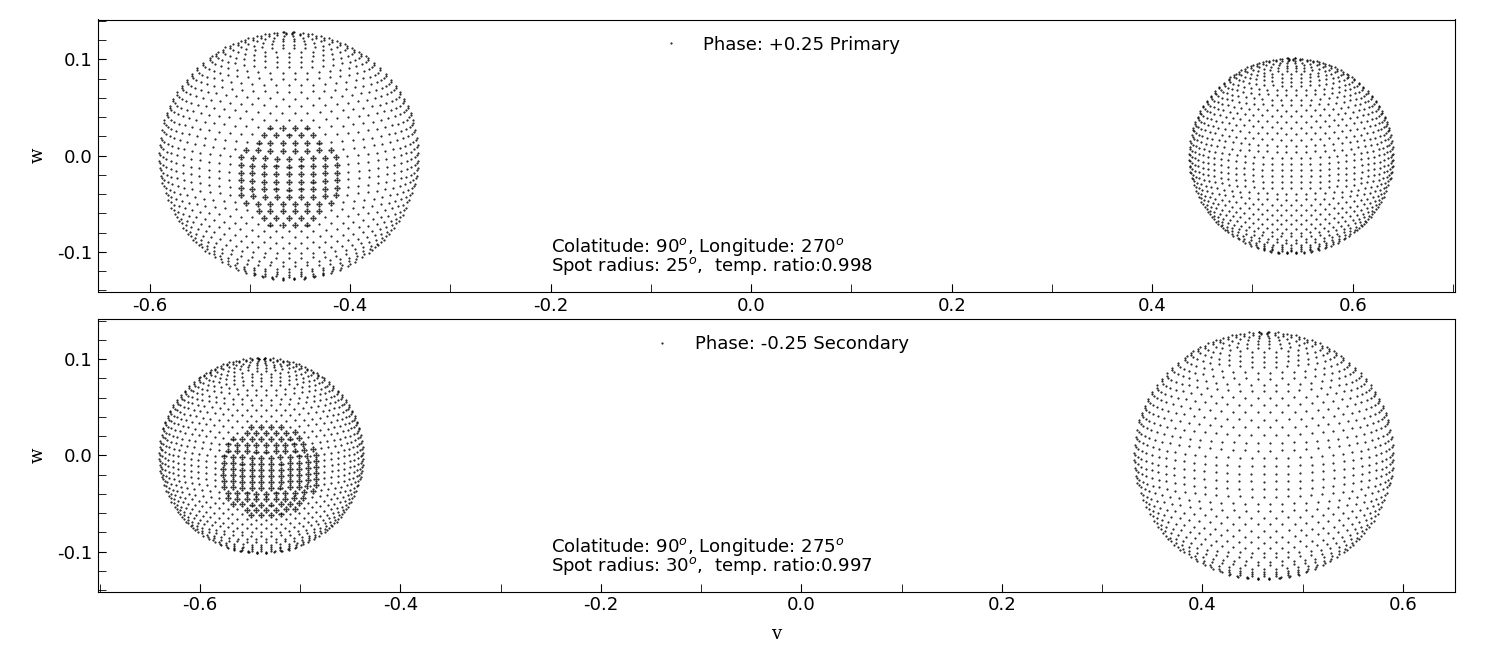}}
\caption{Position of spots on primary (upper panel) and secondary (lower panel) components of the system EPIC5147 as seen during phases 0.25 and -0.25. }
\label{epic_spo}
\end{center}
\end{figure*}

In the \ktwo\ data set, the brightness of system EPIC5147 is found to be slightly different at phases 0.25 and -0.25. The system has slightly high brightness at phase 0.25 as compared to phase -0.25. For this asymmetry of LC, one dark spot at primary and one at secondary are placed. Different positions and parameters of spot are compared on the basis of sum of square of residuals (SSR) to select the final solution.
 The position of a spot on stellar surface is defined by spot colatitude and spot longitude. Spot colatitude is the angular distance of the spot as measured from the North pole of the star (0 to 180$^{\circ}$) while spot longitude is the angular distance of the spot as measured from the direction of the companion in counter clockwise direction (0 to 360$^{\circ}$). The radius of the spot (in degrees) specifies spot size and temperature ratio is defined as the ratio of the spot temperature with respect to local temperature of photosphere. Fig.~\ref{fit_res} shows the residuals of fit with and without the use of spots in the model. The spot parameters are determined as colatitude: 90$^{\circ}$ (fixed), longitude: 270$^{\circ}$ ($\pm$5), spot radius: 25$^{\circ}$ ($\pm$5), and temperature ratio ($T_{spot}/T_{star}$): 0.998 for spot on primary component. For spot on secondary component, colatitude: 90$^{\circ}$ (fixed), longitude: 275$^{\circ}$ ($\pm$5), spot radius: 30$^{\circ}$($\pm$5), and temperature ratio ($T_{spot}/T_{star}$): 0.997. Although the level of asymmetry is very small and actual cause of the uneven brightness is unclear, the inclusion of the spots during the modeling resulted in a slightly better fit. The position of spot is shown in Fig.~\ref{epic_spo}. Both the components are scaled according to a normalized semi-major axis (a). 

\begin{figure*}
\begin{center}
\subfigure{\includegraphics[width=18cm,height=11cm]{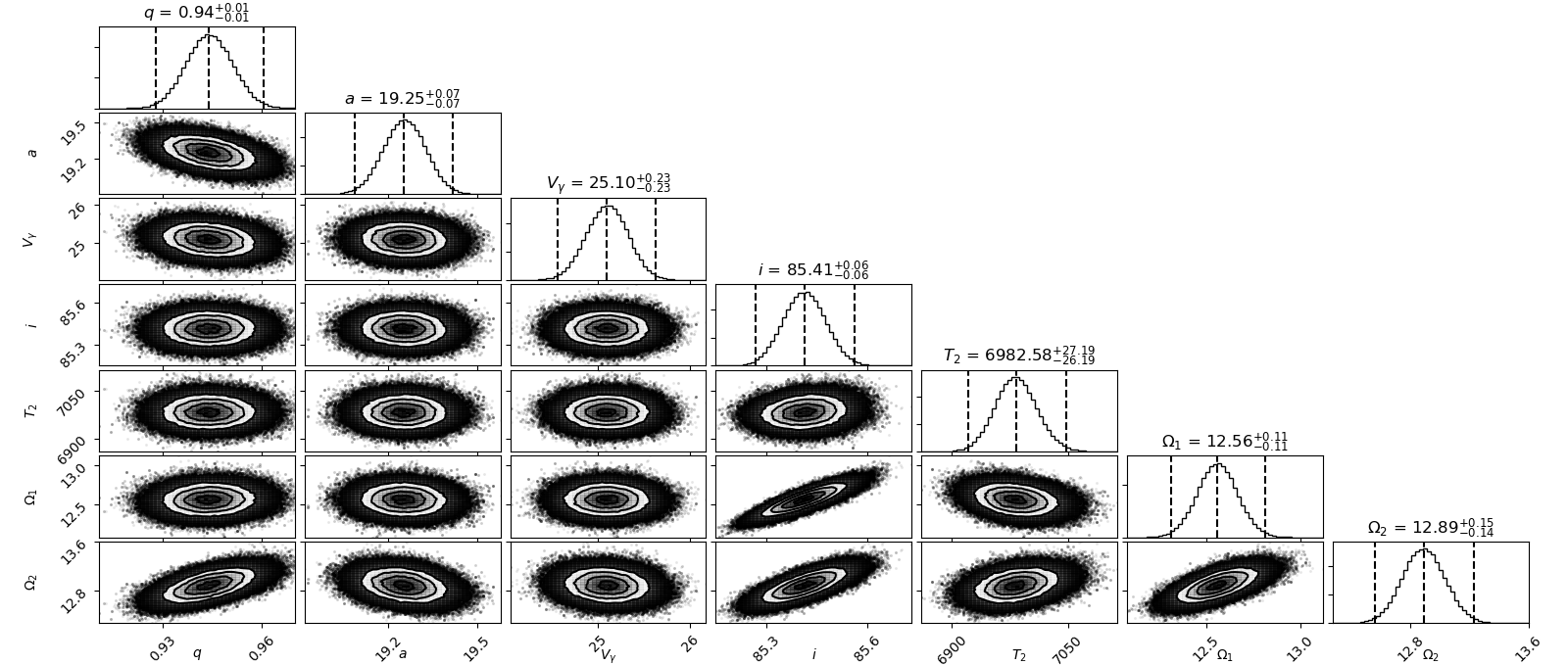}}\vspace{0pt}
\subfigure{\includegraphics[width=18cm,height=11cm]{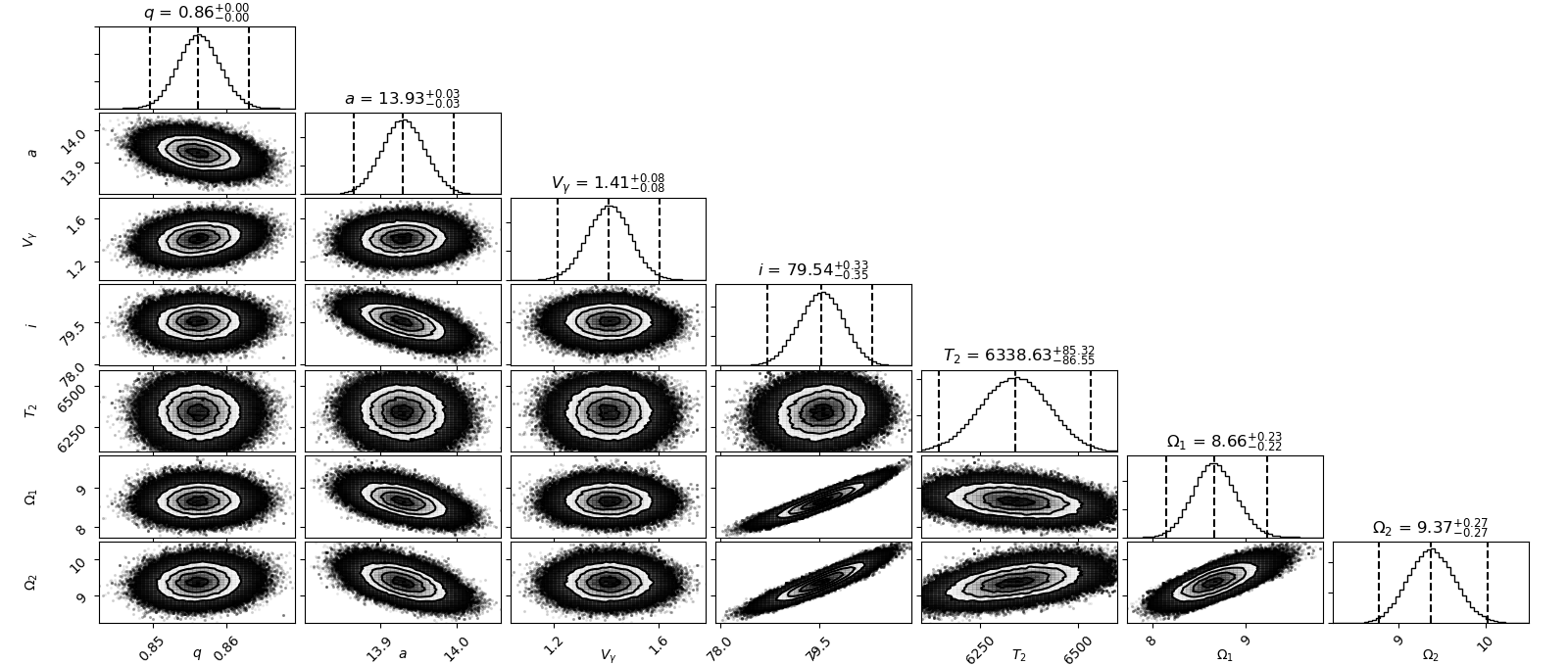}}
\caption{Corner plots MCMC distributions for EPIC2753 (lower panel) and EPIC5147 (upper panel). The vertical dashed lines represent the 1, 50 and 99 \% quantiles of distribution.}
\label{epic1_cor}
\end{center}
\end{figure*}

 The errors given by PHOEBE GUI are fitting errors, which are probably an underestimation of the true errors. To estimate the errors more robustly, we use the Markov Chain Monte Carlo (MCMC) method. A Python script is developed by \cite{2005ApJ...628..426P} which allows the user to apply the EMCEE code \citep{2013PASP..125..306F} with the PHOEBE scripts. The EMCEE code is a Python implementation of Goodman $\&$ Weare’s Affine Invariant Markov chain Monte Carlo (MCMC) Ensemble sampler \citep{2010CAMCS...5...65G}. During the MCMC run, the semi-major axis (a), center of mass velocity ($V_{\gamma}$), mass ratio (q), inclination (i), secondary component's $T_{\rm eff}$ and primary/secondary component surface potential ($\Omega_{1, 2}$) are set as free parameters. Appropriate lower and upper limits are used for these free parameters. A combination of 125 walkers and 8,000 iterations is used for MCMC runs. First 1,00,000 iterations are discarded from all 10,00,000 (125$\times$8,000) iterations as MCMC burn-in period. The corner plots for the MCMC runs are given in Figure~\ref{epic1_cor} for both sources. The quantities at the top of subplots are 1, 50, and 99 percentiles for each distribution. The standard deviation for the MCMC distribution of each parameter is reported as the uncertainty. Table~\ref{mod_para} lists the values for determined parameters from the best fitted model and their MCMC derived uncertainties. The quantities $r_{1, 2}$ in the Table~\ref{mod_para} are the component radii normalized to the semi-major axis or simply the dimensionless/normalized radii. PHOEBE calculates $r_{1, 2}$  using the $\Omega_{1/2}$. Note that the peaks of the MCMC distributions shown in Figure~\ref{epic1_cor} are slightly different from the best model parameters mentioned in Table~\ref{mod_para} as distributions are represented by histograms with some bin width for different parameters.

%

\begin{table}
\caption{The combined LC and RV solutions for both the systems. The '*' indicates the errors are determined using MCMC distribution.}             
\label{mod_para}      
\centering
\begin{tabular}{l l l}    
\hline    
Parameters               & Kepler/\ktwo\  & Kepler/\ktwo\  \\
                         & EPIC2753       & EPIC5147       \\
\hline 
a ($R_{\odot}$)          & 19.26(0.07)*   & 13.92(0.03)*   \\
$V_{\gamma}$ (km/s)      & 25.1(0.2)*     & 1.41(0.08)*    \\
q                        & 0.942(0.007)*  & 0.856(0.003)*  \\
i ($^{\circ}$)           & 85.53(0.05)*   & 79.71(0.34)*   \\
\teffb\                  & 6981(28)*      & 6318(85)*      \\

$l_{1}$                  & 7.414(0.008)   & 8.377(0.007)   \\
$l_{2}$                  & 5.145          & 4.154          \\
$\Omega_{1}$             & 13.56(0.11)    & 8.61(0.23)     \\
$\Omega_{2}$             & 12.88(0.15)    & 9.49(0.27)     \\
$r_{1}$ (a)              & 0.0861(0.0009) & 0.1292(0.0037) \\
$r_{2}$ (a)              & 0.0797(0.0010) & 0.1021(0.0033) \\
\hline                  
\end{tabular}
\vspace{1ex}
\end{table}
\subsection{Absolute Parameters}\label{params}

The orbital solutions derived via model fitting are used to estimate the fundamental parameters such as radius (to make things clear for coming sections, the parameter "radius or $R_{1,2}$" is used to represent the component radius in solar radius units and "normalized radius or $r_{1,2}$" to represent the radius in semi-major axis units), mass and luminosity of individual components. The RV modeling give semi-major axis (a) and with the help of known value of \porb, total mass of the system ($M_{T}=M_{1}+M_{2}$) is calculated. The semi-major axis normalized radii for components are converted to actual radii in solar radius units using the semi-major axis. The mass-ratio q is used to determine the masses of individual components. The radius of each component is derived from model fitting and the temperatures of the individual components are used to estimate the luminosity of the individual components. The total luminosity of the system can be converted to the bolometric luminosity and absolute magnitudes with the help of appropriate bolometric corrections. The distance modulus is related to the absolute and apparent magnitude which give the distance of the system. The parameters and uncertainties are determined using the FORTRAN code JKTABSDIM \citep{2005A&A...429..645S}. JKTABSDIM uses bolometric correction tables by \cite{2002A&A...391..195G} for estimating absolute magnitudes. The absolute parameters are listed in the Table~\ref{abs_param}. The distance estimates from code JKTABSDIM given in Table~\ref{abs_param} make use of the surface brightness relations by \cite{2004A&A...426..297K}.

\subsection{Evolutionary Status}\label{evol_stat}

\begin{figure*}
\begin{center}
\subfigure{\includegraphics[width=18cm,height=8cm]{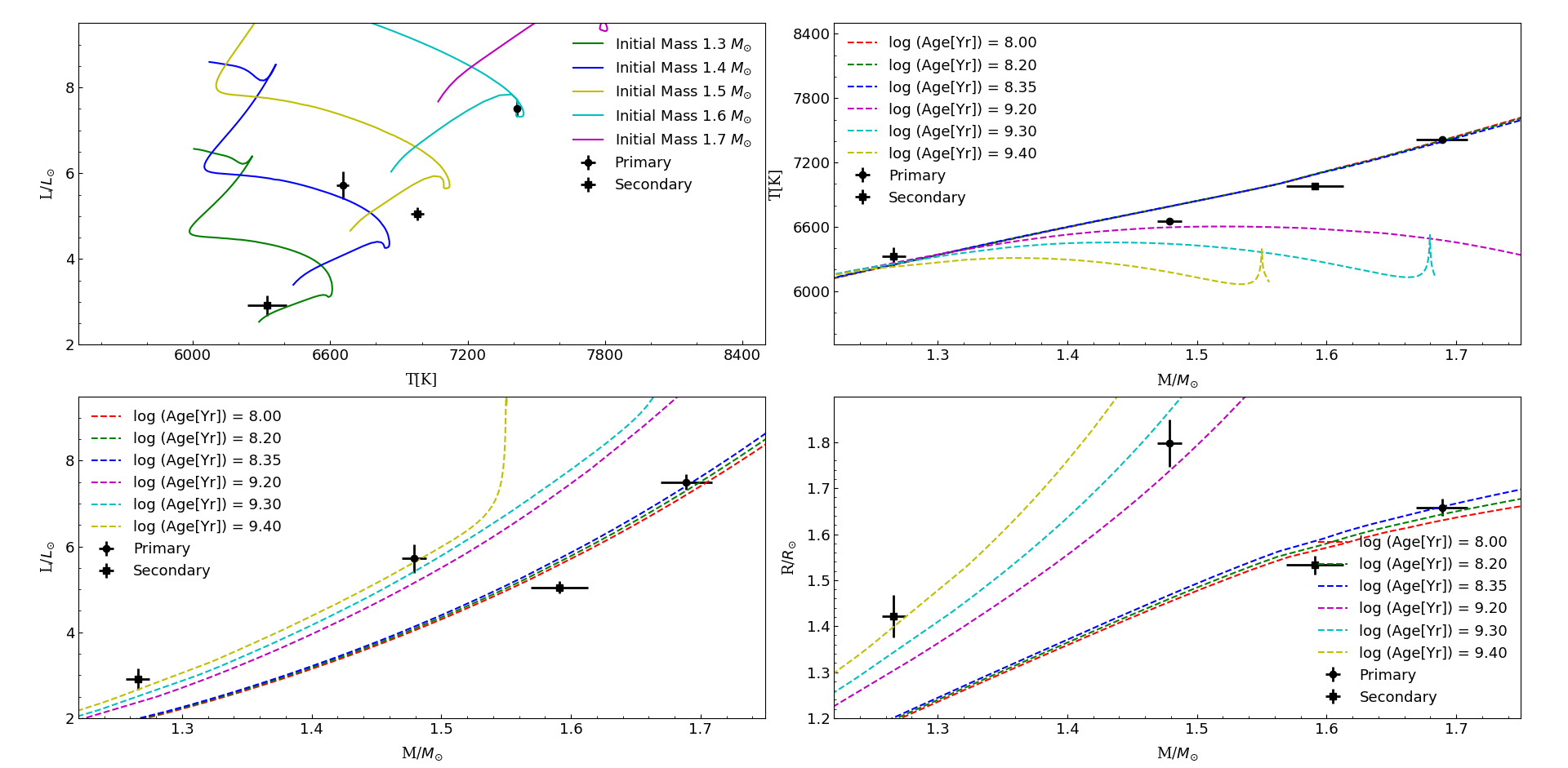}}
\caption{The upper left plot shows the position of components along evolutionary tracks with different initial masses. The other plots show position of targets in M-T, M-L and M-R plane along with different isochrones from MIST.}
\label{evo_epic2}
\end{center}
\end{figure*}

To probe the stellar evolution, isochrones and stellar evolutionary tracks are generated using MESA Isochrones $\&$ Stellar Tracks (MIST\footnote{https://waps.cfa.harvard.edu/MIST/}, \citealt{2011ApJS..192....3P, 2016ApJ...823..102C, 2016ApJS..222....8D}). While generating the stellar evolutionary tracks, the initial masses close to the calculated masses are chosen. For isochrones, the log(Age[Yr]) is varied from 8.00 to 10.00 in steps of 0.05. The metallicity \feh=0.2 is used for stellar evolution tracks/isochrones. Fig.~\ref{evo_epic2} shows the MIST evolutionary tracks/isochrones with the EPIC2753 and EPIC5147 components.  Only the main sequence part of the evolutionary tracks and isochrones is shown in Fig.~\ref{evo_epic2}. Different colors in stellar evolutionary tracks (continuous lines) and isochrones (dashed lines) represents different initial masses and ages, respectively. The components of EPIC2753 lie close to the isochrones with log(Age) = 8.20 to 8.35 (100 Myr-224 Myr). The secondary component of EPIC2753 is found to have slightly low luminosity and radius as expected from the used isochrones. The components of EPIC5147 lies close to the isochrones with log(Age) = 9.20 to 9.40 (1.6-2.5 Gyr).
\begin{table}
\caption{The absolute parameters derived for the system in the present study.}             
\label{abs_param}
\centering          
\begin{tabular}{l l l}
\hline\hline     
Parameters               & Kepler/K2     & Kepler/K2     \\
                         & EPIC2753      & EPIC5147      \\
\hline
d (pc)                   & 238(4)        & 199(5)        \\
$M_{1}$ ($M_{\odot}$)    & 1.69(0.02)    & 1.48(0.01)    \\
$M_{2}$ ($M_{\odot}$)    & 1.59(0.02)    & 1.27(0.01)    \\
$R_{1}$ ($R_{\odot}$)    & 1.66(0.02)    & 1.80(0.05)    \\
$R_{2}$ ($R_{\odot}$)    & 1.53(0.02)    & 1.42(0.05)    \\
$L_{1}$ ($L_{\odot}$)    & 7.50(0.17)    & 5.72(0.33)    \\
$L_{2}$ ($L_{\odot}$)    & 5.05(0.15)    & 2.91(0.24)    \\
\hline                  
\end{tabular}
\vspace{1ex}
\end{table}
%

%

\begin{figure*}
\begin{center}
\subfigure{\includegraphics[width=18cm,height=6cm]{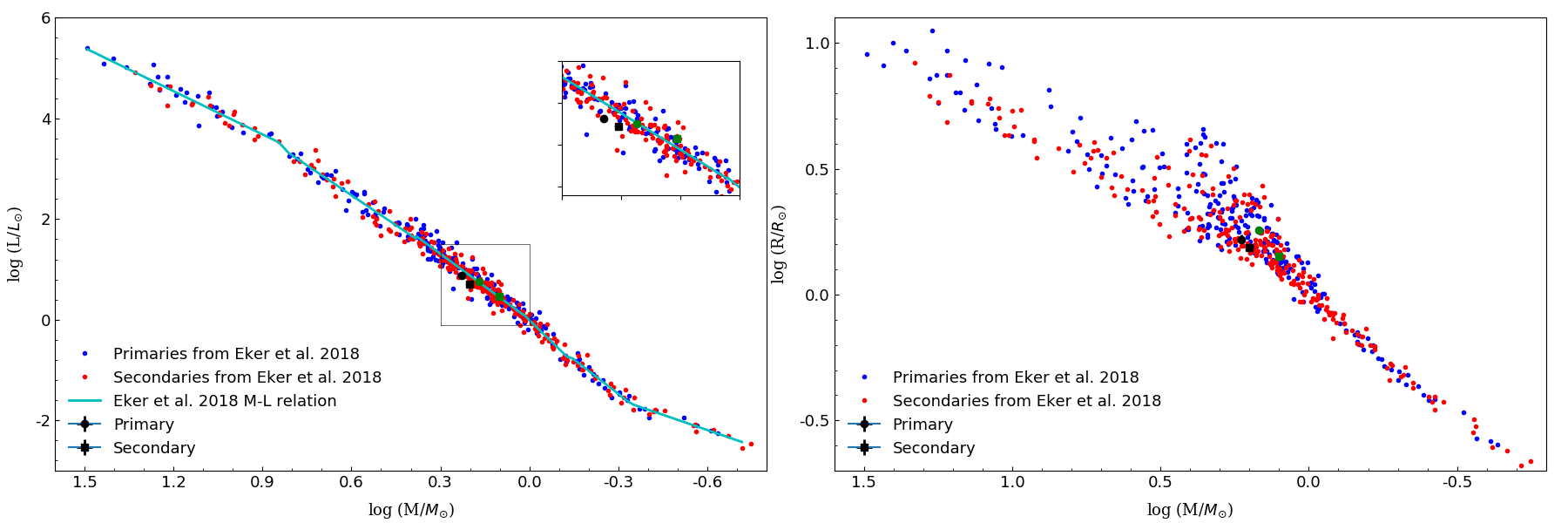}}
\caption{The component parameters along with other well characterized EBs. EPIC2753 and EPIC5147 components are shown in green and black color, respectively.}
\label{eker_epic}
\end{center}
\end{figure*}

\section{Results and Discussion}\label{discu}

Double-lined detached EBs are crucial to test the models of stellar structure and evolution. The multiple campaign photometric data from \ktwo\ is analyzed with the RV data for two eclipsing binary candidates to derive accurate physical parameters. Sources are also observed using the 1.3-m DFOT, Nainital and ASAS-3 survey. The systems are mentioned in some surveys but this work is the first one where a complete characterization of the systems is done. The TOMs are determined using data sets from multiple \ktwo\ campaigns with a time basis of almost 3.2 years. The \porb\ is found to be constant over a period of 3.2 years for both systems. Updated linear ephemeris are derived using these data. The \porb\ is mentioned as 1.8112662 days in \cite{2016AA...594A.100B} for EPIC5147 while we derived \porb\ as 3.6215398 days. The results of \cite{2016AA...594A.100B} are solely based on \ktwo\ C05 observations while the present work gives a more reliable ephemeris for EPIC5147 by including C16, C18 observations. Although present analysis does not highlight any variation in period of these systems, it is impossible to completely discard the possibility of period variation on the basis of a time span as short as 3.2 years.

The masses and radii of the components of EPIC2753 are derived as $M_{1,2}$ = 1.69(0.02), 1.59(0.02) $M_{\odot}$ and $R_{1,2}$ = 1.66(0.02), 1.53(0.02) $R_{\odot}$. Similarly, the masses and radii of the components of EPIC5147 are derived as $M_{1,2}$ = 1.48(0.01), 1.27(0.01) $M_{\odot}$ and $R_{1,2}$ = 1.80(0.05), 1.42(0.05) $R_{\odot}$.

\cite{2018MNRAS.479.5491E} derived mass-luminosity and mass-radius relations using 509 well-studied main-sequence stars in detached eclipsing binary systems. The mass-luminosity data are represented in the form of six-piece classical mass-luminosity relation for six different mass ranges. The mass-radius relation is derived for stellar masses ranging from 0.179 to 1.5 $M_{\odot}$. For EPIC2753 the component luminosities are derived as $L_{1,2}$ = 9.88(0.79) and 7.64(0.65) $L_{\odot}$ using the mass-luminosity relation for intermediate mass stars. For EPIC5147 and $L_{1,2}$ are calculated as 5.56(0.34) $L_{\odot}$ and 2.84(0.16) $L_{\odot}$ from mass-luminosity relation. Similarly, use of mass-radius relation estimated the primary and secondary component radius $R_{1,2}$ as 1.74(0.59) $R_{\odot}$ and 1.39(0.55) $R_{\odot}$ for EPIC5147. For EPIC5147, the results by these relations are same as our estimates within error limits but for EPIC2753, the estimated luminosities are higher than expected values. One reason for this deviation in the results can be the non-homogeneous nature of the sample used by \cite{2018MNRAS.479.5491E}. A large fraction of the stars in the studied sample consists of main sequence stars from the solar neighborhood disc which are mostly metal rich.
 The empirical relations by \cite{2018MNRAS.479.5491E} do not involve the effect of metallicity. Only a limited number of stars in the sample have trustworthy metallicity estimates. The mass-radius and mass-luminosity relations can be different for samples with different ages and metallicity. \cite{2021A&A...647A..90F} derived mass-radius and mass-luminosity relations including the effects of age and metallicity but the results are based on a small sample of 56 stars. The study of their effect on mass-luminosity and mass-radius relations need more large sample of stars with reliable age and metallicity measurements. Figure~\ref{eker_epic} shows the component parameters of studied EBs with other EB components used by \cite{2018MNRAS.479.5491E} to derive the relations. The continuous line in the left panel of Figure~\ref{eker_epic} represents the six-piece M-L relation by \cite{2018MNRAS.479.5491E}.

The distances are derived as 238 (4) and 199 (5) pc for EPIC2753 and  EPIC5147, respectively. The distances for these sources are mentioned as $\sim$ 248 (1) pc (for EPIC2753) and $\sim$ 200 (2) pc (for EPIC5147) in \cite{2021AA...649A...1G} catalog which are close to our estimates. The evolution status of the systems are investigated using the MESA evolutionary tracks and isochrones. The comparison of EPIC2753 components with different isochrones shows that the age of the system is around 100-224 Myr. The age of system EPIC5147 is determined as 1.6-2.5 Gyr on the basis of isochrones. Other constraints such as accurate \logg\ and \feh, can further refine the age estimates.

\section*{Acknowledgements}

This work is supported by the Belgo-Indian Network for Astronomy and astrophysics (BINA), approved by the International Division, Department of Science and Technology (DST, Govt. of India; DST/INT/BELG/P-09/2017) and the Belgian Federal Science Policy Office (BELSPO, Govt. of Belgium; BL/33/IN12). VOSA has been partially updated by using funding from the European Union's Horizon 2020 Research and Innovation Programme, under Grant Agreement nº 776403 (EXOPLANETS-A) . In this work we have also used the data from the European Space Agency (ESA) mission \gaia, processed by the \gaia\ Data Processing and Analysis Consortium (DPAC). This work also make use of the Two Micron All Sky Survey and SIMBAD database.

\section*{Data Availability}

The data supporting the results of this study are available from the corresponding author on reasonable request.



\bibliographystyle{mnras}
\bibliography{main}




%
%


\bsp	
\label{lastpage}
\end{document}